# Building high-energy silicon-containing batteries using off-the-shelf materials

Marco-Tulio F. Rodrigues,[1,z] Stephen E. Trask,[1] Alison R. Dunlop,[1] Yi-Chen Lan,[1] Joseph Kubal,[1] Devashish Salpekar,[1] Andressa Y. R. Prado,[1] Evelyna Wang,[1] Charles McDaniel,[1] Eliot F. Woods,[1] Lily A. Robertson,[1] Ryan J. Tancin,[2] Maxwell C. Schulze,[2] Nicolas Folastre,[1] Baris Key,[1] Zhengcheng Zhang,[1] Wenquan Lu,[1] Daniel P. Abraham,[1] Andrew N. Jansen[1]

[1] Chemical Sciences and Engineering Division, Argonne National Laboratory, Lemont, IL, USA

[2] Energy Conversion and Storage Systems Center, National Renewable Energy Laboratory, Golden, CO, USA

[z]**Contact:** marco@anl.gov

**Abstract**

The technology of silicon anodes appears to be reaching maturity, with high-energy Si cells already in pilot-scale production. However, the performance of these systems can be difficult to replicate in academic settings, making it challenging to translate research findings into solutions that can be implemented by the battery industry. Part of this difficulty arises from the lack of access to engineered Si particles and anodes, as electrode formulations and the materials themselves have become valuable intellectual property for emerging companies. Here, we summarize the efforts by Argonne's Cell Analysis, Modeling, and Prototyping (CAMP) Facility in developing Si-based prototypes made entirely from commercially available materials. We describe the many challenges we encountered when testing high-loading electrodes (> 5 $mAh/cm^2$) and discuss strategies to mitigate them. With the right electrode and electrolyte design, we show that our pouch cells containing ≥ 70 wt% $SiO_x$ can achieve 600-1,000 cycles at C/3 and meet projected energy targets of 700 Wh/L and 350 Wh/kg. These results provide a practical reference for research teams seeking to advance silicon-anode development using accessible materials.

## Introduction

Silicon has a high capacity for $Li^+$ storage and can be used to complement or substitute graphite in the anode of Li-ion batteries. The structural rearrangements needed to accommodate lithium lead to a decrease in the density of the Si host material,[1] causing particle-level dilation that can reach ~300%.[2-4] Repeated cycles of expansion and contraction can produce extensive fracturing of brittle silicon domains,[5] which contributes to electronic disconnection and performance fade. Yet, Si-based materials (be it Si-carbon composites, $SiO_x$, or "pure" silicon nanostructures) are already present in various commercial cells. In many cases, the mitigation strategy for these mechanical issues has been to use these materials sparingly, typically below ~5 wt%,[6-8] so that cells can receive a moderate energy boost with minimal compromise in lifetime. More generous energy gains can be unlocked by designing electrodes where the silicon-based component is the major contributor to anode capacity (> 800 mAh/$g_{Si+C}$; we henceforward refer to these as "Si-rich"). Some U.S. companies are pursuing an aggressive strategy to design high-energy, Si-only, graphite-free cells. This approach offers the potential for energy densities exceeding 750 Wh/L and allows the use of unconventional electrolyte formulations that would normally be incompatible with graphite due to solvent co-intercalation. There is a diversity of strategies being adopted for commercializing silicon-based materials and anodes, and cells benefitting from these approaches are reportedly under pilot-scale production.

One of the remaining challenges for the deployment of Si-rich anodes is calendar aging.[9] In 2020, the U.S. Department of Energy (DOE) published test data on Si-rich cells supplied by undisclosed manufacturers.[9, 10] This data revealed remarkable advancements in the cycle life achieved by these cells, with many surpassing 800 cycles. Nevertheless, all the investigated systems presented unsatisfactory calendar life, exhibiting 20% of capacity fade after less than two years of storage.[9, 10] Although details about testing conditions were not provided, it is assumed that cells were maintained at 100% state of charge (SOC) and above 25 °C. More recent projections based on ongoing tests have hinted at improvements,[11] though disclosure of experimental validation is still pending. Given the importance of calendar life in many applications, the DOE has established the Silicon Consortium Project to investigate the mechanisms of calendar aging and create approaches to reliably expedite life testing and forecasting.

As part of this Consortium, Argonne's Cell Analysis, Modeling, and Prototyping (CAMP) Facility is tasked with developing high-performing electrode formulations and cell designs, and with advancing insights into aging mechanisms through diagnostic analysis. Our prototyping efforts aim to produce surrogates of Si-rich systems of commercial relevance (350 Wh/kg, 700 Wh/L) so that the challenges faced by manufacturers can be replicated and mitigation strategies identified. The CAMP Facility has over a decade of experience working with silicon anodes. Our initial work in this area supported DOE programs aimed at fundamentally investigating the reactivity between silicon and electrolytes. Over time, our emphasis has shifted to cell-level phenomena, evolving alongside Si anodes as they approach market-readiness.

The present work chronicles the CAMP Facility's path in experimenting with Si-rich electrodes, detailing the rationale behind materials selection that enabled 2.3 Ah pouch cells to achieve 1,000 cycles of life. We highlight the challenges encountered, the mitigation strategies adopted, and the lessons learned throughout this process. Importantly, all the experiments described here were conducted using commercially available materials. We hope that this document serves as a practical resource for other research groups aiming to advance Si anode technology.

***Experimental and Modeling***

*Materials sourcing.* Poly(acrylic acid) (PAA, $M_v$ = 450,000 g/mol), lithium hydroxide (LiOH) and N-methyl pyrrolidone (NMP) were sourced from Sigma-Aldrich-Millipore. LiPAA was obtained by titrating aqueous solutions of PAA (15 wt%) with 1 M LiOH until 80% $H^+/Li^+$ substitution was achieved (pH ~6). Polyimide (PI) powder (Mw = 60,000 g/mol) was dissolved (at 20 wt%) in NMP using a rotary mixer for 24 hours at 50-75 RPM; the polymer structure is provided in ref [12]. Polyacrylate binders were acquired from Blue Ocean and Black Stone as aqueous suspensions (Battbond 290S and 290S3). Polyvinylidene difluoride (PVDF, 5130) was purchased from Solvay. Current collectors were battery-grade foils (20 μm aluminum or 10 μm copper). Certain electrodes were coated on a high-tensile-strength Cu-alloy foil (HTA-750, 10 μm), purchased from Schlenk Metal Foils GmbH & Co. Pouch cells were made with a ~150 μm aluminum laminated film (Cell Pack – 153PL, Youlchon Chemical).

Silicon nanoparticles were acquired from various manufacturers, and particle sizes included 30-50 nm or ~200 nm. The $SiO_x$ powder used in much of the work discussed here was acquired from Osaka Titanium Technologies Co. (D50 5.2 μm, 6.3 $m^2/g$) and had no carbon coating. Additional tests included a carbon-coated $SiO_x$ from an alternative manufacturer (D50 5 μm, 2.7 $m^2/g$). NMC811 and NMC532 powders were acquired from Targray and Toda, respectively. Single-walled carbon nanotube (SWNT) suspensions were acquired from Tuball and contained either PVDF or carboxymethyl cellulose (CMC) as stabilizer depending on the solvent (NMP or water, respectively). C-45 carbon black was purchased from Timcal.

All electrolytes used in this work were based on mixtures of organic carbonates with $LiPF_6$. Our baseline electrolyte was a 1.2 M solution of $LiPF_6$ in 3:7 wt EC:EMC (Gen2, Tomiyama), with 3 wt% fluroethylene carbonate (FEC, Solvay) as additive; EC is ethylene carbonate and EMC is ethylmethyl carbonate. For additional electrolyte formulations, vinylene carbonate (VC), dimethyl carbonate (DMC), and diethyl carbonate (DEC) were acquired from Sigma-Aldrich-Millipore, and EMC from Gotion. The separator was Celgard 2500. All materials were used as received unless otherwise specified.

*Electrode fabrication and processing.* Electrode slurries were made with either NMP (for PVDF and PI binders) or water (for LiPAA and the polyacrylates) as solvents. These slurries typically presented a solid content of ~30 wt% (for Si nanoparticles and $SiO_x$) and 55-75% for NMCs (depending on constituents, composition, and loading). Details about mixing procedures and hardware can be found in ref. [13]. Electrodes were cast using a reverse comma coater (A-Pro Co., Ltd.) at a rate of 0.5 m/minute. The coater is equipped with two one-meter-long drying zones that were set at 80-85/105 °C for NMP and 85/105 °C for aqueous slurries. Electrodes were calendered to the specified porosity using a roll press (A-Pro) at 0.5 m/s, at 80 °C for PVDF and ~25 °C for the remaining binders.

The NMC811 cathode contained 96 wt% of active material, 1.95 wt% C-45, 0.05 wt% SWNT, and 2 wt% PVDF. NMC532 laminates comprised 90 wt% active material, 5 wt% C-45, and 5 wt% PVDF. Both were calendered to ~33% porosity. Anode compositions are provided in the text.

*Laser ablation.* The electrodes were patterned using a solid-state, diode-pumped ultrafast laser (Advanced Optowave FEMTO-IR-1030) producing 600 fs pulses centered at ~1030 nm. A dual axis galvo scanner (Aerotech, AVG10HPO) and a telecentric, fused-silica f-theta lens (LINOS, 70 mm focal length) focused the laser to a ~37 µm spot size, which was scanned at 100 mm/s across the electrode surface in a hexagonal pattern with 1 mm spacing between parallel hexagon edges. A 15 µm buffer was used to prevent intersecting line segments from ablating at vertex points repeatedly. The electrode pattern was ablated in two passes of the laser to the full depth of the electrode using pulse energies of 28 μJ (2.6 J/cm$^2$) and 14 μJ (1.3 J/cm$^2$), respectively. Damage to the current collector was minimized by using a lower pulse energy for the 2$^{nd}$ pulse.

*Cell assembly and testing.* All experiments were carried out at 30 °C. Testing was performed in coin cells (2032 format) or pouch cells, both assembled in a dry room. Coin cells used 16 mm separator, and 14 mm positive and 15 mm negative electrodes. For pouch cells, positive/negative electrodes were 14.1/14.9 cm$^2$, or 46.3/49.1 cm$^2$, depending on the format (xx3450 and xx6395, respectively); exact dimensions for the latter are provided in Figure 9. For pouch cells, electrodes were punched from inner areas of the electrode laminates to ensure thickness uniformity, and the coatings were carefully removed from the tabs to expose the current collector. Electrodes were vacuum-dried overnight at 120 °C prior to cell assembly.

For pouch cells, nickel and aluminum tabs were welded to the anode and cathode, respectively, using MWX100 ultrasonic welder (Branson). The pouch material was heat-sealed at 175 °C (tab) or 170 °C (sides). "Dry cells" were dried at 60 °C overnight prior to electrolyte filling. Vacuum sealing was performed on a FCB-200 Fuji Impulse sealer at 164 °C and -85 kPa, which was always followed by additional pressing at the side-area heat-sealer. Forming of the pouch material, tab trimming, and heat sealing were performed on equipment supplied by Media Tech Co. Multi-layer pouch cells were built with the assistance of a semi-automated Z-fold stacking machine acquired from that same company. The electrolyte quantity added to all cells was calculated as a multiple of the total pore volume of electrodes and separator; this pore volume factor is provided in the figure captions when relevant. Pouch cells were tested under a stack pressure of 26 psi, which is similar to the experienced by electrodes in coin cells.[14]

$SiO_x$ full-cells were built using prelithiated anodes. Prelithiation was performed electrochemically by first assembling a cell containing the anode and a suitable Li source. In most tests, this source was Li metal, but multi-layer pouch cells employed NMC811. The anode in this initial cell experienced lithiation to ~50 mV vs $Li/Li^+$, and then delithiation to ≥ 0.95 V vs $Li/Li^+$; an exception were the electrodes from our initial $SiO_x$/PI electrodes, that were delithiated to 0.6 V vs $Li/Li^+$ prior to disassembly (see Figure 3 and Figure 4). This procedure supplies enough $Li^+$ to counter the initial irreversibility of $SiO_x$ and leaves the anode sufficiently "empty" to accommodate the surplus $Li^+$ delivered by the fresh NMC811 cathode when harvested anodes are used in a new, reconstructed full-cell. The prelithiated anodes were dipped in DMC and dried prior to reuse. Prelithiation is discussed in more detailed later in the text. Prelithiation was always performed with the same electrolyte used in the reconstructed cells. Protocols for aging of the reconstructed cells are provided in the text.

We tested electrodes with a variety of compositions, including formulations with or without carbon black, carbon nanotubes, or graphite. Because all these materials contribute reversible capacity within voltage ranges of interest, their weight cannot be disregarded when reporting specific capacities. Capacities are therefore reported in units of $mAh/g_{Si+C}$, where Si denotes the weight of silicon nanoparticles or $SiO_x$ in the electrode coating.

*Additional characterization.* Electrode-level dilation was evaluated on an ECD-3-nano electrochemical dilatometer (EL-Cell GmbH) using 8-mm anode discs, and 1 M LiFSI in 3:7 wt EC:EMC (Soulbrain) as the electrolyte. A detailed discussion about test conditions and cycling protocol is provided in ref. [3].

For nuclear magnetic resonance (NMR) spectroscopy, samples were harvested from cycled cells and packed into 1.3-mm zirconia rotors inside an argon-filled glovebox (< 0.1 ppm $O_2$ and <

0.5 ppm $H_2O$). Rotors were spun at a magic-angle-spinning (MAS) rate of 50 kHz. $^7Li$ NMR measurements were performed on a 300 MHz (7.04 T) Bruker Avance III spectrometer using a 1.3-mm HXY WB MAS probe. Single-pulse excitation sequences were used with a 15 second pulse delay.

Scanning electron microscopy (SEM) was performed on a Jeol IT800HL using an acceleration voltage of 30 kV and a probe current of 75 mA. $SiO_x$ powder was dispersed on a piece of carbon tape adhered to a metal stub.

*BatPaC modeling.* Extrapolation of energy metrics was performed using Argonne's Battery Performance and Cost (BatPaC) techno-economic model. The model considered stiff pouch cells in which the anode had the same specific capacity, areal loading, and porosity exhibited by electrodes made and tested by CAMP (as specified later in the text). The average voltage of the cell was also the value measured in our prototypes. The assumed cathode was a state-of-the-art NMC811 with 96/2/2 wt% composition of active material, binder, and carbon additive, respectively, with 25% porosity and exhibiting 214 mAh/g after formation. The simulated cells contained a 15-µm polyolefin separator (50% porosity), and electrodes were modeled as being coated on 15-µm aluminum and 10-µm copper current collectors. Assumed densities were 2.3 and 4.65 g/cm$^3$ for the anode and cathode active material, respectively, and 1.2 g/cm$^3$ for the electrolyte. The electrolyte content amounted to 1.69-1.91 g/Ah, depending on cell size. Cells with capacities ranging from ~3.6 to ~184 Ah were considered. Cell thickness was fixed at 12 mm, while the lateral dimensions were adjusted to match the target capacity, maintaining a length-to-width ratio for the cathode of ~12. Additional details can be found in the BatPaC documentation.[15]

***Results and Discussion***

***$SiO_x$ as active material.*** The choice of active material is arguably the most important variable in formulating silicon-based electrodes. The high surface area and reactivity of certain Si materials make them susceptible to oxidation by the solvent in aqueous slurries, releasing flammable hydrogen gas.[16, 17] Furthermore, the dimensional changes experienced by silicon during cycling often necessitate mitigation strategies at the particle and/or electrode level to achieve stable operation.[18-21] Thus, additional considerations are required when introducing silicon into the anode, both in terms of processing and performance.

Argonne's CAMP Facility aims to support sustained research efforts to advance Li-ion battery technology. To fulfill this mission, the materials we use must ideally exhibit three key features: i) be suited for roll-to-roll coating techniques; ii) be available in kilogram scale; iii) be free of disclosure restrictions. Satisfying these requirements ensures that our resources can be leveraged to supply consistent electrodes to dozens of researchers during multiyear programs and that our findings can be made available to the battery community. The relative recency of Si-rich anodes in pilot-scale commercial prototypes makes it difficult to access the most advanced materials as companies understandably safeguard their intellectual property.

With these constraints in mind, we have experimented with many types of Si nanoparticles over the years, with sizes ranging from 30-50 nm to ~200 nm. Some of these particles contained surface coatings,[16, 22] though none exhibited structural modifications (such as yolk-shell designs or internal porosity).[19, 23] Figure 1a summarizes historical data obtained from full-cells pairing NMC532 with various anode formulations, using Gen2 + 10 wt% FEC as the electrolyte; all anodes contained LiPAA as the binder. The active material was either a blend of Si and graphite (with 15 or 30 wt% Si), or pure silicon (constituting 70 or 80 wt% of the coating). Regardless of the type and quantity of silicon, cells always displayed rapid capacity fade, primarily due to loss of $Li^+$ inventory from persistent growth of the solid electrolyte interphase (SEI). However, many of these systems also exhibited some amount of anode capacity loss,[4, 24, 25] signaling that particle fracturing remained a problem.[5] The use of alternative electrolytes did not substantially change the outcome.[26]

The observation of Si capacity loss in many of the cells in Figure 1a suggested that building high-capacity electrodes with these nanoparticles would be challenging. Given the lack of openly available commercial "pure Si" alternatives exhibiting lower volume change, we turned to $SiO_x$ (Osaka Titanium Technologies Co.). $SiO_x$ particles consist of silicon nanodomains embedded within a matrix of $SiO_2$. While this arrangement decreases the specific capacity, it also lowers the overall particle dilation and helps shield the active domains from the electrolyte.[18, 27] On the downside, the abundant $SiO_2$ domains react irreversibly with $Li^+$ to form a variety of silicates, decreasing the initial coulombic efficiency of the electrode. Many commercial cells using "Si-graphite" anodes actually contain $SiO_x$,[6, 8] generally in limited amounts to minimize this initial irreversibility. Deploying anodes with a high fraction of $SiO_x$ requires prelithiation to achieve energy gains.

The simple move to $SiO_x$ drastically changed the dimensional behavior of the electrodes during cycling. Figure 1b compares the thickness change of electrodes containing 70 wt% of either a Si nanoparticle (blue) or $SiO_x$ (orange) as they are lithiated to 10 mV vs. Li/$Li^+$. Using $SiO_x$ led to only ~20% of the expansion observed with Si (~59 vs ~270%) but delivered > 60% of the capacity. Although limiting the Si electrode to ~1,200 mAh/g would result in a maximum dilation similar to that observed with $SiO_x$, that would also lead to a higher average delithiation potential,[4] lowering the cell-level energy gains. Furthermore, our previous attempts at constraining the Si utilization in full-cells resulted in limited gains in cyclic stability,[4] suggesting that detrimental changes in particle integrity and morphology continue to take place in the "pure Si" electrodes.[5] Attempting to replicate the capacity achieved with $SiO_x$ using a Si-graphite blend (Figure 1b, gray) required 30 wt% silicon and nonetheless led to higher expansion (~109%).

As we discuss below, adoption of $SiO_x$ markedly improved cycle life. Additionally, the composite nature of its micron-sized particles (Figure 1c) decreases the surface area of active material in direct contact with the electrolyte, which may benefit calendar life. However, this change in active material also lowers the specific capacity of electrodes. We next examine the implications of this lower capacity for our goal of achieving cells with 350 Wh/kg and 700 Wh/L.

***Techno-economic considerations.*** We used Argonne's BatPaC model to evaluate how changing the properties of the anode would affect the energy metrics of the cell. The anode was assumed to contain 87 wt% active material, with several possible values considered for specific and areal capacity, as well as porosity. The cathode was NMC811, and the ~73 Ah pouch cells had an average voltage of ~3.43 V at C/3. For simplicity, the initial irreversibility of all electrodes remained constant and equaled that of BatPaC's standard graphite electrode. Refer to the *Experimental and Modeling* section for additional details about some of the assumptions used here. Importantly, the specific capacity refers to the *accessible* capacity of the anode as Si-containing electrodes can be deployed with a variety of effective voltage windows.[4] The effects of Si volume expansion and prelithiation on energy metrics were neglected here, leading to somewhat optimistic projections; these effects are discussed in a later section.

Figure 2a and Figure 2b show the energy density and specific energy, respectively, that can be achieved according to the specific capacity of the active material of cells with a 3.5 $mAh/cm^2$ anode with 35% porosity. Our targets of 350 Wh/kg and 700 Wh/L are, in principle, within reach of materials like $SiO_x$ (1,000-1,200 $mAh/g_{Si+C}$) and apparently with sufficient margin such that prelithiation and changes in electrode design can later be accommodated. Notably, meeting the volumetric energy density target appears easier than reaching 350 Wh/kg.

Graphs like Figure 2a and Figure 2b are common in the literature and are often used to support the conclusion that capacities > 1,000 $mAh/g_{Si+C}$ are superfluous, as the rate of energy gain decreases sharply beyond that point while electrode expansion does not (Figure 1b). Nevertheless, such representations omit an important practical point, which is how thick the electrode needs to be to reach a given energy target. Figure 2c displays the accessible areal capacities that the anode must exhibit at each assumed specific capacity for ~73 Ah pouch cells to achieve 350 Wh/kg. The different traces represent cases with different assumed anode porosities. Clearly, the required areal capacity rises sharply as the specific capacity decreases: while 1,500 $mAh/g_{Si+C}$ can enable 350 Wh/kg already at ~3.5 $mAh/cm^2$ and 55% porosity, decreasing the specific capacity to 1,000 $mAh/g_{Si+C}$ raises the areal requirements to ~5 $mAh/cm^2$. This latter loading is similar to the exhibited by graphite anodes in certain commercial Li-ion batteries with particularly thick electrodes.[28] Silicon anodes are generally considered suited for fast charging due to their elevated average lithiation potentials (thus making Li plating less likely) and their thinner designs.[2, 29, 30] Figure 2c shows that this latter advantage diminishes somewhat as specific capacities decrease.

We used Figure 2c as a guide for electrode design. For $SiO_x$ electrodes (1,000-1,200 $mAh/g_{Si+C}$), meeting our energy targets would require 3.3-5 $mAh/cm^2$ anodes with 35-55% porosities (Figure 2c). An appropriate binder is needed to maintain electrode cohesion in the presence of volume change at such high loadings.

***Aromatic polyimide binder.*** Much research has focused on developing binders endowed with structural features specifically tailored for silicon. Carboxymethyl cellulose, sodium alginate, polyacrylic acid, and PAA derivatives are among the most investigated materials[31-36] as they can bind to surface groups at the Si particles to ensure the formation of lasting interfaces. It has also been argued that these polymers can tune the SEI composition by regulating the accessibility of species to the active material surface.[37] All these polymers have been tested at the CAMP Facility over the years with various silicon nanoparticles,[38] and LiPAA (pH ~6) was generally found to offer the best balance between electrode performance and slurry processability with the materials we used. The extent of $Li^+/H^+$ exchange in LiPAA is regulated by the pH and affects the surface charges along the polymer chains, which influence their conformation and their mixing with Si particles.[36] Given our historical experience with LiPAA, this polymer was initially among the binders we tested with $SiO_x$. Electrodes with the composition 70 wt% $SiO_x$, 10 wt% Timcal C45 carbon, and 20 wt% LiPAA displayed promising initial performance (Figure 3a), exhibiting ~1,230 $mAh/g_{Si+C}$ (~1.5 $mAh/cm^2$) of capacity at C/10 and 1,000 $mAh/g_{Si+C}$ at 2C in half-cells operating between 0.05 and 1.5 V.

Although encouraging, these initial tests with LiPAA were performed at much lower loadings than required to approach our energy targets (Figure 2c). The need for thick electrodes led to the consideration of an aromatic polyimide (PI) as an alternative binder. Besides having excellent adhesive properties,[39-41] this polymer can be an electronic conductor,[12, 42] potentially improving accessibility to active material in high-loading electrodes. Preferential lithiation of anode particles closer to the separator has been reported in graphite cells at moderate and fast rates.[43, 44] In Si electrodes, such heterogeneous lithiation could have serious consequences due to the resulting stress concentrations.

Initial electrodes containing the PI binder displayed adhesion to the Cu foil superior to that obtained with LiPAA as indicated by peel tests (not shown here). Furthermore, experiments in half-cells showed that replacing LiPAA with PI increased electrode capacity by ~200 mAh/g at all rates at a similar mass loading (Figure 3a). Structural integrity also seemed to improve with the polyimide. Figure 3b compares the dQ/dV traces for the LiPAA electrode, after five cycles at C/10, and after enduring over ~130 cycles (0.05 – 1.5 V) at various rates; in the latter case, the cell was rebuilt by harvesting the electrode and pairing it with fresh Li metal and electrolyte. All features related to the reversible redox of Si presented significantly lower amplitude after aging, indicative of permanent deactivation of certain electrode domains. Similar experiments done with the PI-based electrode (Figure 3c) showed better retention of features, indicating that the accessibility to $SiO_x$ particles was better preserved.

This initial evidence supported the decision to adopt the polyimide as binder. Full-cell testing in coin cells at slightly higher loadings (~2.2 $mAh/cm^2$ of utilized capacity vs NMC811, 3.0-4.06 V) resulted in ~1,000 cycles of life (Figure 3d), a significant improvement from the data in Figure 1a.

***Issues with the polyimide binder.*** Notwithstanding the favorable performance achieved with the PI binder, we identified a few drawbacks with implementation of this polymer. The most obvious one relates to processability. Polyimides are traditionally made from the reaction between a dianhydride and a diamine, forming a polyamic acid that is dehydrated at high temperature during imidization.[40, 41] The aromatic PI we used here had the advantage of being pre-imidized and, in principle, would dispense such thermal treatment. We found, however, that annealing the electrodes at 350 $^{o}$C for one hour significantly improved capacity retention during cell testing. Interestingly, spectroscopic analyses revealed no structural changes in the polymer during this process, suggesting that improvements were likely related to subtle increases in the cross-linking density. Despite the improved performance, this additional step would increase time and cost for large-scale anode production. Moreover, this PI is insoluble in water and requires NMP for processing, going against recent industry trends of transitioning to lower-cost aqueous slurries.

A second difficulty we encountered with the aromatic PI related to its reactivity. We showed in Figure 3a that $SiO_x$ electrodes using the PI binder outperformed the LiPAA ones by ~200 mAh/$g_{Si+C}$ at all rates. We later identified that this difference was not due to an improved utilization of the $SiO_x$ particles but rather to a partially reversible redox activity of the binder itself. This is better illustrated in Figure 4a, which shows that delithiation of $SiO_x$ proceeds identically with both binders until ~0.7 V, when $Li^+$ extraction from the polyimide commences. In a separate detailed study,[12] we have demonstrated that this polyimide can have an initial $Li^+$ uptake of > 2,000 mAh/$g_{PI}$ and present sustained reversible capacities that are comparable with that of graphite (> 300 mAh/$g_{PI}$). However, we also showed that this reactivity appears to accelerate the calendar aging and could lead to underestimation of overall cell degradation.[12]

A final challenge with working with this aromatic polyimide was that, paradoxically, we found it to be too adhesive. After prelithiation, we would observe the electrodes to have irreversibly stretched laterally by as much as ~6% due to plastic deformation of the copper current collector.[39] Furthermore, both the coating and Cu exhibited visible ripples (Figure 4b). These effects are a consequence of the adhesion strength of the binder. When $SiO_x$ particles expand during the initial lithiation, the lack of interfacial debonding leads to all the stress being transferred to the Cu foil. At sufficiently high $SiO_x$ content and coating loading, the resulting forces overcome the elastic limits of copper, causing plastic deformation (Figure 5a). During delithiation, contraction of $SiO_x$ particles causes an inward pull on the foil, which could result in the formation of wrinkles (Figure 5b). We have previously shown that this rippling only occurs for electrodes larger than a critical length, being observed in our pouch cells (≥ 14.9 $cm^2$) but not in coin cells (≤ 1.77 $cm^2$).[39] Formation of these ripples will damage the electrode coating and favor the occurrence of Li plating (Figure 4b), penalizing the performance of larger cell formats (Figure 4c).[39] Additional tests (not shown) indicated that neither the rippling nor the resulting Li plating could be prevented by increasing the stack pressure of the cells from 4 to 72 psi (~0.03 to ~0.5 MPa), confirming that the observations arose from lateral dilation/contraction of the coating film.

Naturally, the selective occurrence of these rippling in large electrodes was a major barrier in our pursuit of demonstrating long cycle life in ~2 Ah pouch cells. Since the problem is caused by the strong adhesive properties of the PI, a solution would involve weakening the bonding of the coating to the current collector, with minor localized interfacial failure being a favorable compromise to rippling (Figure 5c). This outcome could possibly be achieved by lowering the binder content in our electrodes (initially 20 wt%, from our prior experience working with Si nanoparticles). However, we considered that its redox activity and the need for thermal treatment made this aromatic PI an unsuitable option, prompting us to explore alternative binders to eliminate wrinkling.

***Implementing a polyacrylate copolymer binder.*** Our experiments with the PI binder indicated a way forward: identify a polymer with slightly lower adhesion to copper that also allows for aqueous processing. The former would be achieved not only by finding a suitable binder but also by decreasing the overall binder content from the 20 wt% used initially. Given that the stresses within the electrodes we had tested already appeared to be high, formulations with a $SiO_x$ content higher than 70 wt% were initially not pursued. We also made the decision to eliminate carbon black from the electrode in an attempt to minimize the conductive surface area and thus decrease calendar aging. Instead, graphite was introduced as a conductive additive, which is also expected to help improve electrode capacity. Because the micron-sized graphite flakes lack the percolating ability of carbon black nanoclusters, electronic connectivity was further supplemented with single-walled carbon nanotubes (SWNTs).

Our team considered multiple binder options and electrode compositions, which will not be discussed here for brevity. Our exploration eventually led us to the following electrode formulation: 70 wt% $SiO_x$ (Osaka), 19 wt% MagE3 graphite (Hitachi), 1 wt% SWNT (Tuball), 1wt% CMC, and 9 wt% Battbond 290S polyacrylate (Blue Ocean and Black Stone), with a calendered porosity of ~57%; CMC was present as stabilizer in the aqueous SWNT dispersion. Besides presenting desirable mechanical properties, the selected polyacrylate binder was sourced directly from a manufacturer serving the battery industry, which in our experience tends to increase batch-to-batch reproducibility of the product.

With this new formulation, we succeeded in building $SiO_x$ electrodes with loadings > 5 $mAh/cm^2$ (0.05 – 0.7 V) that did not exhibit rippling after the initial cycling. Figure 6a displays the cycle life of a 46.3 $cm^2$ single-layer pouch cell pairing this electrode with NMC811 and using Gen2 + 3 wt% FEC as electrolyte. The anode was initially prelithiated vs Li metal, with final delithiation to 1.05 V vs $Li/Li^+$. The cell was tested within the voltage range of 4.17-2.8 V and presented an initial C/10 capacity of 5.58 $mAh/cm^2$ (992 $mAh/g_{Si+C}$). The pouch cell maintained a C/10 capacity retention > 80% for 665 cycles, after which it exhibited rollover failure (Figure 6a). Teardown of the cell after testing revealed no evidence of Li plating or electrode wrinkling (Figure 6b,c), suggesting that failure was likely caused by electrolyte depletion. Evidently, this composition appears capable of supporting extended cycling of electrodes with high $SiO_x$ content and the high areal loadings required to achieve 350 Wh/kg (Figure 2b).

Despite this encouraging start, this electrode still required improvements. Due to unreliable access to more scalable approaches at the time of this work, our prelithiation process involved assembling a cell containing the anode vs a suitable Li source (which could be lithium metal or NMC811) and performing at least one full cycle. The anode would then be harvested from this original pouch cell, dipped in DMC, and paired with a fresh cathode to constitute the full-cell that would be aged. With the electrode formulation of Figure 6, we occasionally observed minor delamination of the prelithiated anode during the initial disassembly process, with flakes of coating along the edges remaining attached to the separator (Figure 7a), signaling opportunities for improvement. We later identified that this partial delamination was prompted by the lateral expansion of the coating, which would partially debond from copper and extend beyond the rim of the current collector (Figure 8b). This observation is a striking example of cohesion within the coating surpassing the adhesion strength with copper in the presence of large internal stresses. A more adhesive binder, such as the PI, would have pulled the copper along, producing the ripples seen in Figure 4b and Figure 8a. Although the stresses generated by the coating are about the same, the fraction transferred to the current collector differs, leading to a different outcome.

Given the initial success of this electrode composition (Figure 6a), we investigated whether additional calendering could mitigate this delamination. Encouragingly, half-cell tests indicated that decreasing the porosity from ~57.2 to ~41% did not visibly affect performance (Figure 7b). However, disassembly of coin cells revealed that lower porosities caused complete delamination of the coating (Figure 7c). We have previously shown that higher internal porosity can accommodate the initial expansion of Si particles during lithiation, decreasing the bulk electrode dilation.[3] Here, the choice and quantity of binder led to a formulation unable to cope with the additional expansion experienced at lower porosities, causing the coating to be peeled away with the separator during disassembly. This observation suggests that an optimal electrode should have a somewhat higher binder content and porosities > 50%. Naturally, high porosities will necessitate high areal loadings to achieve our energy targets (Figure 2c).

Further testing revealed that replacing 3 wt% of graphite with binder (resulting in 16 wt% graphite and 12 wt% polyacrylate in the formulation) produced electrodes that consistently maintained their cohesion after prelithiation and avoided plastic deformation of the Cu current collector (Figure 8c; 5.2 $mAh/cm^2$ and 950 $mAh/g_{Si+C}$, 0.05-0.7 V vs Li). However, issues with plastic deformation of copper reappeared when testing a double-sided version of this electrode (Figure 8d). Certain mechanical behaviors exhibited by single-sided electrodes (such as curling) tend to be absent from double-sided coatings as effects on each side cancel out each other.[45, 46] Here, since wrinkling is caused by lateral forces imposed by dilation and contraction of $SiO_x$,[39] adding a second layer of coating effectively doubles the resulting stress, creating mechanical issues where they appeared to have been mitigated. This observation was an important lesson that would later inform the experimental screening of additional electrodes. Despite the rippled electrodes being prone to Li plating (Figure 4b), they were nonetheless used to assemble multi-layer pouch cells.

***2 Ah cell design.*** Our multi-layer pouch cells used 46.3 $cm^2$ NMC811 cathode and 49.1 $cm^2$ $SiO_x$ -graphite anodes (Figure 9a). Since the anodes presented 5.2 $mAh/cm^2$ per side of coating, a total of 10 interfaces (that is, five double-sided anode punches) were needed to surpass 2 Ah. Given the dimensional changes experienced by $SiO_x$ particles during cycling, we designed all anode layers to be double-sided to ensure a constant mechanical environment within the cell. The resulting 10 interfaces were paired with four double-sided cathodes and two additional single-sided layers at the cell extremes (Figure 9b).

As discussed above, conversion of silica domains into silicates lowers the initial coulombic efficiency of $SiO_x$. For our final anode formulation, efficiency was only ~54% when cycled between 0.05 and 0.7 V; part of that loss is only apparent, due to capacity that is accessible only at higher electrode potentials.[47] Given this low value, the electrode requires prelithiation to enable cells with high energy. In our electrochemical approach, prelithiation essentially comprises building another cell to perform the initial formation cycle. At a ~54% initial efficiency and 5.2 $mAh/cm^2$ reversible anode capacity, the first lithiation of the anode would require almost 10 $mAh/cm^2$ of Li. Using lithium metal as counter-electrode would produce very reactive surfaces; furthermore, it would also lead to chemical reduction of the electrolyte, and some of the resulting soluble species could persist in the prelithiated anode after disassembly. Thus, safety and performance considerations guided our choice of using NMC811 (~12 $mAh/cm^2$, 2.5 – 4.3 V) as a counter-electrode for prelithiation.

The multi-layer pouch cells built for prelithiation contained 49.1 $cm^2$ cathode and anode layers. Since a perfect replication of the electrode alignment is difficult to achieve, using equally-sized electrodes helped prevent the existence of an overhang of unreacted $SiO_x$, which could later consume large amounts of $Li^+$ in the reconstructed full-cell. After filling the cells with the desired electrolyte and vacuum-sealing them, they were placed in a testing fixture under 26 psi and moved to an environmental chamber held at 30 ºC. The cells rested for 30 minutes for thermal equilibration and then experienced a tap charge to 1.5 V followed by 6 hours of rest; this helps drive the anode away from Cu-corrosion potentials and enables a longer wetting period prior to cycling.[48] The tap charge and rest steps were repeated three additional times for a total rest time of 24+ hours. Prelithiation consisted of C/10 cycling with capacity-limited charge and discharge based on knowledge of the anode half-cells. Afterwards, cells were disassembled, and the anodes were harvested and dipped in DMC. The dried, prelithiated anodes were then used to assemble a reconstructed cell vs lower-loading fresh NMC811 for aging (Figure 9c). As shown in Figure 8d, the harvested anodes presented ripples on the coating and current collector.

Following prelithiation, the anodes had an open circuit potential of ~0.95 V vs. $Li/Li^+$. This relatively high potential helped ensure that the reconstructed cell was not “cathode-limited” at the end of discharge.[49] After the first charge of NMC electrodes, some of the extracted $Li^+$ is kinetically inhibited from returning to the cathode lattice, generating some initial irreversibility.[50] If this excess $Li^+$ is not consumed in SEI-forming reactions (as in the case of a prelithiated anode), the surplus will remain at the anode and form a reservoir that can delay side-reactions from leading to measurable capacity fade.[49] Although that can help extend cell life, it also makes it more difficult to evaluate how experimental variables are affecting aging outcomes. In our cells, the anode was sufficiently emptied of $Li^+$ at the end of prelithiation so that this surplus from the cathode can be accommodated, avoiding the formation of this reservoir.

Prelithiation could also have been done by performing a partial lithiation of the anode instead of a full cycle with excess delithiation. The partially-lithiated anodes would then be harvested and used to reconstruct the cells. This would be somewhat akin to depositing lithium metal directly onto the anode as is being pursued by some research groups and commercial entities.[51-53] Nevertheless, we reasoned that residual electrolyte and exposure to dry room air could lead to self-discharge and thus poorer control of the state of lithiation of these electrodes during cell reconstruction. (Note that scalable Li deposition approaches are usually "dry" processes.) On the other hand, our chosen approach has two caveats. First, exposing the anode to high potentials could damage the SEI,[54, 55] decreasing the effectivity of prelithiation. Second, excessive delithiation of $SiO_x$ could increase the compressive stresses that cause wrinkling and induce mechanical damage to the particles themselves.[56, 57] Despite these challenges, this method was found to yield highly reproducible results.

***Testing multi-layer pouch cells with the baseline electrolyte.*** 2 Ah pouch cells were tested with three different electrolytes. In this section, we will focus on experiments performed using two cells containing the baseline system (Gen2 + 3 wt% FEC), to illustrate basic properties and discuss some key highlights.

The reconstructed cells were tested within the voltage window of 4.18 and 2.8 V and were first exposed to eight cycles at C/10. At the end of these conditioning cycles, cells delivered 2.34 Ah (Figure 10a), reaching 5.05 mAh/cm$^2$ and 926 mAh/$g_{Si+C}$ per interface. Differential voltage analysis (Figure 10b) revealed that the cathode and anode reached ~4.25 / 0.07 V vs $Li/Li^+$ at the end of charge, and ~3.55 / ~0.75 V vs $Li/Li^+$ at the end of discharge. Cells then underwent rate tests, comprising initial and final cycles at C/10, and three cycles each at C/5, C/3, and 1C. Notwithstanding the high areal loading of the electrodes, high capacities were retained at all rates, with only a ~4% decrease at C/3, and a ~9% drop at 1C (Figure 10c). Initial cell impedance was very consistent across both cells, being ~36 Ω $cm^2$ at low SOCs and decreasing to ~21 Ω $cm^2$ as charging proceeded (Figure 10d). This higher resistivity at low states of lithiation is typical for silicon-based materials.[25]

After this initial characterization, cells were aged at C/3, with reference performance tests (RPTs) being carried out every 100 cycles. RPTs consisted of a full cycle at C/10 and pulse-based resistance measurements (10 second pulses at 2C for discharge and 1.5C for charge; see ref. [48] for details about the technique). A constant voltage step at the end of discharge was performed prior to each RPT until the current dropped below C/10. For cell 1, RPTs were performed every 50 cycles to increase data resolution.

During aging, cells exhibited linear capacity fade during the initial ~400 cycles (~110 days of testing) but consistently experienced rollover failure afterwards, prompting the end of life (i.e., 20% fade) to be reached at ~461 cycles (Figure 10e). The rollover also involved a discontinuity in the impedance rise (Figure 10f), which otherwise presented a slow but steady increase over time. The aging trends observed for these cells were similar to the shown in Figure 4c, where life was limited by wrinkling-induced Li plating. Plated lithium is reactive and can eventually lead to local electrolyte dry-out,[58] explaining the impedance rise we observed (Figure 10f). Injection of additional fresh electrolyte to the cell (along with vacuum sealing) restored some of the capacity but did not affect the fade rate (Figure 10g). The impedance, however, decreased sizably after electrolyte addition (Figure h). These observations are consistent with the idea of continuous Li plating (the rate of which is unaffected by the electrolyte volume in the cell) dictating capacity loss and with electrolyte depletion producing impedance rise. At the end of testing, cells exhibited visible gassing (Figure 10i), and whitish deposits around the wrinkled surfaces confirmed the occurrence of plating (Figure 10j). This preferential nucleation of Li on the deformed ridges of the current collector echoes our previous observations.[39]

***Multi-layer pouch cells with alternative electrolytes.*** Multi-layer pouch cells, built as described above, were also tested using two additional electrolytes: 1.2 M $LiPF_6$ in EC/EMC/DEC 20/40/40 vol + 3 wt% FEC + 1 wt% VC; and 1.2 M $LiPF_6$ in EMC/VC 10/2 wt; this latter composition has been recently suggested to yield a stable SEI in silicon.[59]

Using the electrolyte containing the three carbonate solvents led to identical capacity fade profiles to the baseline electrolyte (Figure 11a, orange vs gray markers), suggesting that the aging was again dictated by the occurrence of Li plating. Clearly, in this case the changes in electrolyte formulation had no visible effect on the fate of the cell. The VC-rich composition, however, showed monotonic losses during the entire test, remarkably reaching a cycle life of ~1,000 (Figure 11a, red) after ~260 days. While a welcome result, this cyclic stability was also puzzling as it hinted at the first example of a cell containing a rippled anode that did not exhibit Li plating. This was further suggested by the lack of a sharp increase in impedance rise at the end of life (Figure 11b), suggesting that electrolyte dry-out was also avoided in this system. Indeed, the cell experienced negligible gassing after testing (Figure 11c), in stark contrast to the exhibited by the other electrolytes (Figure 11d and Figure 10i). Interestingly, disassembly of the cells revealed that all of them experienced Li plating, including the one with the VC-rich electrolyte (Figure 11e).

The ability of this electrolyte to support stable cycling of the $SiO_x$ electrode, without significant reactivity toward Li metal, is intriguing and is the subject of ongoing investigation. For now, it demonstrates that, with the right electrolyte, this electrode formulation can achieve long cycle life if wrinkling (or its consequences) is prevented.

As an additional note, the initial impedance decrease in the cell with the VC-rich electrolyte was found to be quite reproducible. We suspect that this behavior may be related to the presence of graphite in our electrodes as VC is known to form an effective, but resistive, surface layer on this material.[60] Our team is currently exploring alternative formulations that provide similar functionality without compromising initial cell impedance.

***Extrapolating energy metrics.*** Our cell design in Figure 9 did not attempt to minimize the weight and volume of inactive components, such as tabs, current collectors, separator, and the pouch material itself. Consequently, directly calculating specific energy and energy density from the physical properties of the resulting cells would fall short of evaluating our final anode composition. Instead, we used BatPaC to determine the energy metrics of an ideal cell containing our $SiO_x$ electrode and otherwise employing state-of-the-art components and built in automotive-relevant pouch cell formats.

It was assumed that the anode had the same composition, utilized capacity (926 $mAh/g_{Si+C}$) and porosity (57%) exhibited by the electrode in Figures 9-11 and that the modeled cell presented the same C/3 average voltage observed in our prototypes (3.43 V). From half-cell data, it was determined that the initial irreversibility of the anode would require the equivalent of a 11-μm compact layer of Li metal to be deposited over the electrode during prelithiation; this Li would spontaneously react with $SiO_x$ prior to cell assembly,[52] contributing weight but not volume to the electrode. We further considered that the modeled cell had a N/P ratio of 1 (as the utilized rather than total anode capacity was provided as input), resulting in a cathode loading of 5.05 mAh/cm$^2$ after formation (the same as exhibited by our cells). Additional assumptions for the NMC811 cathode and other components are provided in the *Experimental and Modeling* section.

The resulting energy density (Wh/L) and specific energy (Wh/kg) are shown in Figure 12 as a function of the assumed cell capacity. Both energy metrics exhibited rapid initial rise with increasing cell sizes, generally plateauing above ~50 Ah. This important observation shows that benefits from Si-based anodes to pouch cells are mostly accrued at larger cell capacities. Our target 700 Wh/L is reached in cells larger than ~33 Ah (Figure 12a, blue trace), with > 750 Wh/L being achievable above 70 Ah.

Nevertheless, these estimates do not account for volume changes occasioned by $SiO_x$ expansion. We also considered cases in which the cell is designed to accommodate this dilation, by taking the effective anode thickness to be the one it would exhibit at the end of cell charge. Dilatometry experiments (not shown here) with our final electrode formulation revealed a maximum thickness change of ~30% when the anode is lithiated to 70 mV vs. $Li/Li^+$ (Figure 10b). BatPaC modeling incorporating this 30% expansion predicted lower energy densities than in the previous case (Figure 12, orange), with the 700 Wh/L target being achieved at ~100 Ah. These latter estimates are likely conservative as the dilatometry experiments were performed under a relatively low stack pressure (2.9 psi, or ~20 kPa); electrodes can experience significantly higher pressures in operating cells,[61, 62] which can constrain electrode dilation.[63-65]

Figure 12b shows that our model cells would display ~340 Wh/kg at a ~73 Ah capacity, and that accounting for $SiO_x$ volume expansion has little effect on this metric. Although this specific energy is slightly below our initial target of 350 Wh/kg, it could be improved by increasing the accessible capacity of the electrode, as discussed in a later section.

All in all, despite the high initial porosity, electrodes containing 70 wt% of $SiO_x$ can successfully approach our target for specific energy and surpass the one for energy density. This is only possible with prelithiation. Otherwise, the initial inefficiency of this material would necessitate a cathode with loading much higher than the required by the reversible capacity of the anode, resulting in poorer energy metrics (Figure 12a,b, gray traces).

***Calendar aging tests.*** One of the core objectives of our work developing Si-dominant prototypes is to enable the study of calendar aging in systems that can approach the performance requirements of commercial cells. The discussion above shows that our final electrode formulation meets that expectation. The challenge that persists with these high-loading electrodes is wrinkling of the current collector as the resulting Li plating could affect aging trends with most electrolytes. To address this issue, preliminary calendar aging studies were conducted using single-layer pouch cells, where the absence of a backside coating prevents electrode wrinkling (Figure 8c).

The calendar aging protocol is based on recommendations by the United States Advanced Battery Consortium (USABC, ref. [66]), ensuring that our tests are compatible with over two decades' worth of data acquired with commercial cells. In the protocol, RPTs were spaced by month-long periods of open circuit storage at 100% SOC (Figure 13a). Once a day, cells experienced pulses (30 second discharge at 1C and 10 second charge at 0.75 C) to provide high-resolution impedance and relaxation information. These pulses were always followed by a residual recharge to 4.18 V (i.e., 100% SOC) with a hold for five minutes to minimize SOC drift due to self-discharge. RPTs consisted of three full cycles at C/10, followed by pulse-based impedance measurements (with same rate and duration as the daily pulses).

Cells using the baseline electrolyte (Gen2 + 3 wt% FEC) have so far been on test at 30 ºC for over a year and a half. Capacity fade is progressing notably slowly, with only ~4.3% lost after 18 months (Figure 13b). The rate of fade diverges from the $\sqrt{t}$ ($t$ being the time at storage) commonly reported in some of the calendar aging literature.[67, 68] Impedance rise is also progressing at a relatively mild pace, with a ~17% increase experienced after maintaining the cells for an extended period at ~100% SOC (Figure 13c).

These favorable results are in contrast with the unsatisfactory calendar life previously reported for high-energy Si-based prototypes.[9, 10] A key distinction between such cells and ours is the electrolyte content: 4.4x in excess of the total pore volume in our case, vs assumedly lean conditions in the pilot-scale commercial cells of refs. [9, 10]. Although our tests can provide some indication about the expected rate of SEI growth,[47] other relevant aging mechanisms (such as certain sources of impedance rise and loss of active material) may only become visible as the availability of electrolyte wanes.[69, 70] We believe that the excellent calendar life of our cells is partly due to the absence of such mechanisms, supporting our hypothesis that time-dependent aging of Si cells is dictated by electrolyte depletion.[24, 47] The cell failure due to dry-out in Figure 6a (also in the absence of Li plating) at much shorter test times (~7.5 months) suggests that cycling contributes to aging much more aggressively than idle storage.

Long-term testing of cells with lower electrolyte content and alternative electrolyte formulations are ongoing and will be discussed in future publications.

***Mitigating current collector deformation.*** The results discussed above show that the cyclic stresses imposed by $SiO_x$ combined with the good adhesion of the binder to the current collector can lead to plastic deformation and rippling of both coating and copper (Figure 5). This deformation, in turn, favors the occurrence of Li plating, which can limit cell performance. Tuning the adhesion strength to the Cu foil can be a countermeasure to this issue (Figure 8c), though not always an effective one (Figure 8b and Figure 8d). Thus, to consistently eliminate rippling, we could: i) use a stronger current collector; ii) modify the electrode architecture to better accommodate the $SiO_x$ expansion; and iii) change the active material. We have tried all the above.

Doping of copper can lead to large increases in tensile strength, which in principle could mitigate the plastic deformation that precedes wrinkling. Our team has successfully used copper alloys with various silicon materials, and some of our exploits with these alternative current collectors are summarized elsewhere.[71] In our experience, however, using these alloys will not always mitigate the mechanical issues experienced by the electrodes. In cases where the active material exhibits excessive dilation, or where the Si content and loading are sufficiently high, stresses from lithiation can overcome the adhesion between the electrode coating and the foil. In these cases, the foil will remain pristine, but the coating can partially delaminate or totally debond from the current collector.

If selection of the active material is constrained, modifying the electrode architecture is also an option. The strategic addition of porosity to the electrode through laser ablation has previously been shown to alleviate mechanical issues in Si-based cells.[20, 21] We used this approach to carve hexagonal trenches 1 mm apart in our double-sided $SiO_x$ electrodes, with the same formulation that previously underwent wrinkling after prelithiation (Figure 8d). The honeycomb pattern of the electrode was maintained after prelithiation and, importantly, no deformation of the current collector was observed (Figure 14a). This prelithiated anode was then tested in a ~0.47 Ah bilayer pouch cell, containing two single-sided NMC811 cathodes (not ablated), and Gen2 + 3 wt% FEC as electrolyte. Although non-negligible, the overall amount of material ablated was small enough to result in no visible difference in initial performance compared to the original ~2.3 Ah pouch cells (Figure 14b). The aging behavior, however, was distinct: rollover failure was still observed but only after ~600 cycles (Figure 14c). Unlike the previous cells, however, the laser ablated electrode did not exhibit Li plating as deposits were not visible (Figure 14d) nor detectable through $^{7}$Li NMR of scraped areas of the coating (not shown), suggesting that rollover may be due to cell dry-out (like in Figure 6a). Even without the life-limiting effects of Li plating, the baseline electrolyte was unable to achieve longer cycle lives.

As mentioned above, an additional approach to mitigate wrinkling is to use active materials that expand less and thus generate lower stresses. Next, we discuss our preliminary studies with other active materials.

***Reaching for higher energies.*** We have demonstrated that, by changing the electrode architecture and using a suitable electrolyte, electrodes containing 70 wt% of $SiO_x$ can be used to build high-performing cells that would approach ~686 Wh/L and ~339 Wh/kg in automotive-relevant pouch cell formats (including prelithiation and accommodating electrode expansion). How can we reach even higher energies?

One option would be to fabricate electrodes with higher areal loading, but the > 5 mAh/cm$^2$ of our anodes is already comparable with the upper bound currently found in commercial cells.[28] An alternative would be to decrease the initial porosity of the electrodes, though that would possibly magnify mechanical issues (Figure 7c). Another option would be to increase the $SiO_x$ content in the electrodes. We have explored this latter approach and found it impractical with the material we use. The wrinkling of the double-sided electrodes in Figure 8d indicates that, at the high areal loadings required for $SiO_x$ to deliver high energy, we were already close to the maximum stresses that the system can handle. We attempted compositions with 75-86 wt% of $SiO_x$ using high-tensile-strength copper alloys as current collectors and consistently observed single-sided electrodes delaminating upon disassembly after a few cycles. This indicates that the stresses from $SiO_x$ cycling are exceeding the binding energy of the coating with the foil. Many of these compositions were viable at lower loadings that were insufficient to reach our energy targets (Figure 2c). Increasing the active material content would require exploring systems that generate less stress.

We were recently able to source a $SiO_x$ powder from an alternative vendor, and preliminary results are encouraging. Initial trials indicated that a blend with the Osaka $SiO_x$ provided great balance between specific capacity and dimensional stability, resulting in the following formulation: 70 wt% $SiO_x$ (Manufacturer A), 18 wt% $SiO_x$ (Osaka), 1 wt% SWNT (Tuball), 1 wt% CMC, and 10 wt% Battbond 290S3 polyacrylate (Blue Ocean and Black Stone), coated on a copper alloy (HTA-750, Schlenk Metal Foils GmbH & Co.). The new $SiO_x$ powder has slightly lower capacity and appears to undergo less volume change than the material sourced from Osaka. The properties of $SiO_x$ depend on the $Si/SiO_2$ ratio, which may explain these differences. Combining the two flavors of $SiO_x$ enabled us to completely remove graphite from the electrode while retaining the active material content (that is, $SiO_x$ + carbons). Despite the higher overall $SiO_x$ content, a double-sided electrode did not exhibit wrinkling or delamination after prelithiation (Figure 15a). Bilayer cells fabricated vs NMC811 (Figure 15b; ~5.15 mAh/cm$^2$ and ~1075

mAh/$g_{Si+C}$) and using the baseline electrolyte endured ~700 cycles before experiencing rollover failure. Once again, Li plating was not visible (Figure 15d) nor detectable through NMR, suggesting that life could be further extended by improved electrolyte formulations.

Using these electrode and cell properties in BatPaC, we estimate that a ~73 Ah pouch cell could reach 809 Wh/L and 353 Wh/kg if electrode expansion is ignored. When accounting for the same 30% expansion used in Figure 12, the projected values decrease to 730 Wh/L and 352 Wh/kg, still exceeding our original targets.

## *Conclusions*

Silicon anodes have come a long way, from a brittle impossibility to a marketable product. Despite the use of Si-rich anodes in pilot-scale commercial cells, replicating their long cycle life in academic settings remains challenging. This difficulty arises, in part, from limitations in sourcing the most advanced materials as companies in this field continue to mature. The undesired consequence is that academic research is often done in systems with sub-par performance and/or scale, raising questions about the transferability of their conclusions to the issues faced by the battery industry.

This manuscript describes in some detail the efforts by Argonne's CAMP Facility to overcome this limitation. We reviewed our historic work with silicon nanoparticles and used dilatometry experiments to justify the pivot towards $SiO_x$. We also detailed our experience with an aromatic polyimide binder that displayed promising initial performance but ended up being impractical. Some of our main concerns with this PI binder are the need of a high-temperature annealing, its reactivity that appeared to accelerate calendar aging, and its excessive adhesive strength that causes plastic deformation of the current collector. Many of these problems were solved by transitioning to a polyacrylate binder, with the added benefit of it being compatible with aqueous processing.

A series of 2.3 Ah multi-layer pouch cells were built using our final $SiO_x$ -rich formulation. Remarkably, tests using an EC-free electrolyte resulted in 1,000 cycles of life with no visible gassing. BatPaC modeling showed that large-format pouch cells using our anode would reach ~340 Wh/kg and >700 Wh/L if built with capacity higher than ~70 Ah, indicating that our design successfully approached our energy targets.

Notwithstanding the performance of our formulation, it still led to plastic deformation of the Cu foil in double-sided electrodes. We have demonstrated two solutions for this issue. One was to improve the anode architecture using laser ablation, which assisted in relieving the stresses arising from expansion of $SiO_x$ particles. The second was to modify the electrode constitution by implementing blends of two commercial $SiO_x$ powders and coating the slurry on a high-tensile-strength Cu alloy. These latter electrodes achieved > 700 cycles of life with the baseline Gen2 + 3 wt% FEC electrolyte, with projected energy metrics of 730 Wh/L and 352 Wh/kg at a ~72 Ah scale, accounting for both prelithiation and volume expansion.

All the work presented here relies on materials that are commercially and openly available at scale. We hope this report serves as a practical resource for academic teams as they develop high-quality prototypes for use in silicon anode research.

***Acknowledgements***

This research was supported by the U.S. Department of Energy's Vehicle Technologies Office under the Silicon Consortium Project, directed by Brian Cunningham, Thomas Do, Nicolas Eidson and Carine Steinway, and managed by Anthony Burrell. The submitted manuscript has been created by UChicago Argonne, LLC, Operator of Argonne National Laboratory ("Argonne"). Argonne, a U.S. Department of Energy Office of Science laboratory, is operated under Contract No. DE-AC02-06CH11357. 

***References***

1. H. Kim, C.-Y. Chou, J. G. Ekerdt, and G. S. Hwang, *The Journal of Physical Chemistry C*, **115**, 2514 (2011).

2. M. Flügel, M. Bolsinger, M. Marinaro, V. Knoblauch, M. Hölzle, M. Wohlfahrt-Mehrens, and T. Waldmann, *Journal of The Electrochemical Society*, **170**, 060536 (2023).

3. A. Y. R. Prado, M.-T. F. Rodrigues, S. E. Trask, L. Shaw, and D. P. Abraham, *Journal of The Electrochemical Society*, **167**, 160551 (2020).

4. M.-T. F. Rodrigues, A. Y. R. Prado, S. E. Trask, S. Ahmed, A. N. Jansen, and D. P. Abraham, *Journal of Power Sources*, **477**, 229029 (2020).

5. S. Pidaparthy, M. Luo, M.-T. F. Rodrigues, J.-M. Zuo, and D. P. Abraham, *ACS Applied Materials & Interfaces*, **14**, 38660 (2022).

6. A. Zülke, Y. Li, P. Keil, R. Burrell, S. Belaisch, M. Nagarathinam, M. P. Mercer, and H. E. Hoster, *Batteries & Supercaps*, **4**, 934 (2021).

7. I. Laresgoiti, H. Yi, D. Koster, F. Karimi, J. Yang, D. Schulte, and E. Figgemeier, *Journal of Energy Storage*, **98**, 112918 (2024).

8. F. Alcaide, G. Álvarez, E. Bekaert, F. Bonilla, E. Gucciardi, I. Urdampilleta, R. Vicedo, and E. Ayerbe, *Journal of The Electrochemical Society*, **170**, 080523 (2023).

9. J. D. McBrayer, M.-T. F. Rodrigues, M. C. Schulze, D. P. Abraham, C. A. Apblett, I. Bloom, G. M. Carroll, A. M. Colclasure, C. Fang, K. L. Harrison, G. Liu, S. D. Minteer, N. R. Neale, G. M. Veith, C. S. Johnson, J. T. Vaughey, A. K. Burrell, and B. Cunningham, *Nature Energy*, **6**, 866 (2021).

10. B. Cunningham, Silicon and Intermetallic Anode Portfolio Strategy Overview, in *Annual Merit Review (US Department of Energy, 2020)*.

11. B. Cunningham, Silicon and Intermetallic Anode Portfolio Strategy Overview, in *Annual Merit Review (US Department of Energy, 2023).*

12. S. Rajendran, H. Liu, S. E. Trask, B. Key, A. N. Jansen, and M.-T. F. Rodrigues, *Journal of Power Sources*, **583**, 233584 (2023).

13. S. E. Trask, Y. Li, J. J. Kubal, M. Bettge, B. J. Polzin, Y. Zhu, A. N. Jansen, and D. P. Abraham, *Journal of Power Sources*, **259**, 233 (2014).

14. J. S. Okasinski, I. A. Shkrob, A. Chuang, M.-T. F. Rodrigues, A. Raj, D. W. Dees, and D. P. Abraham, *Physical Chemistry Chemical Physics*, **22**, 21977 (2020).

15. K. W. K. Knehr, Joseph J.; Nelson, Paul A.; Ahmed, Shabbir, Battery Performance and Cost Modeling for Electric-Drive Vehicles (A Manual for BatPaC v5.0), in (2022).

16. M.-T. F. Rodrigues, S. E. Trask, I. A. Shkrob, and D. P. Abraham, *Journal of Power Sources*, **395**, 289 (2018).

17. K. A. Hays, B. Key, J. Li, D. L. Wood, III, and G. M. Veith, *The Journal of Physical Chemistry C*, **122**, 9746 (2018).

18. T. Chen, J. Wu, Q. Zhang, and X. Su, *Journal of Power Sources*, **363**, 126 (2017).

19. N. Liu, H. Wu, M. T. McDowell, Y. Yao, C. Wang, and Y. Cui, *Nano Letters*, **12**, 3315 (2012).

20. J. S. Pope, Maxwell C.; Tancin, Ryan J.; Singh, Avtar; Verma, Ankit; Preimesberger, Juliane; Coyle, Jaclyn; Colclasure, Andrew; Ban, Chunmei; Tremolet de Villers, Bertrand; Finegan, Donal, *SSRN* (2025).

21. Y. Zheng, H. J. Seifert, H. Shi, Y. Zhang, C. Kübel, and W. Pfleging, *Electrochimica Acta*, **317**, 502 (2019).

22. R. T. Haasch, S. E. Trask, M.-T. F. Rodrigues, and D. P. Abraham, *Surface Science Spectra*, **27** (2020).

23. H. Jia, J. Zheng, J. Song, L. Luo, R. Yi, L. Estevez, W. Zhao, R. Patel, X. Li, and J.-G. Zhang, *Nano Energy*, **50**, 589 (2018).

24. M.-T. F. Rodrigues, Z. Yang, S. E. Trask, A. R. Dunlop, M. Kim, F. Dogan, B. Key, I. Bloom, D. P. Abraham, and A. N. Jansen, *Journal of Power Sources*, **565**, 232894 (2023).

25. M. Klett, J. A. Gilbert, S. E. Trask, B. J. Polzin, A. N. Jansen, D. W. Dees, and D. P. Abraham, *Journal of The Electrochemical Society*, **163**, A875 (2016).

26. N. M. Johnson, Z. Yang, Q. Liu, and Z. Zhang, *Journal of The Electrochemical Society*, **169**, 040527 (2022).

27. J. Asenbauer, T. Eisenmann, M. Kuenzel, A. Kazzazi, Z. Chen, and D. Bresser, *Sustainable Energy & Fuels*, **4**, 5387 (2020).

28. M. Ank, A. Sommer, K. Abo Gamra, J. Schöberl, M. Leeb, J. Schachtl, N. Streidel, S. Stock, M. Schreiber, P. Bilfinger, C. Allgäuer, P. Rosner, J. Hagemeister, M. Rößle, R. Daub, and M. Lienkamp, *Journal of The Electrochemical Society*, **170**, 120536 (2023).

29. S. Friedrich, J. Mährlein, A. Durdel, and A. Jossen, *Journal of The Electrochemical Society*, **172**, 070518 (2025).

30. Z. Yang, S. E. Trask, X. Wu, and B. J. Ingram, *Batteries*, **9**, 138 (2023).

31. B. Lestriez, S. Bahri, I. Sandu, L. Roué, and D. Guyomard, *Electrochemistry Communications*, **9**, 2801 (2007).

32. I. Kovalenko, B. Zdyrko, A. Magasinski, B. Hertzberg, Z. Milicev, R. Burtovyy, I. Luzinov, and G. Yushin, *Science*, **334**, 75 (2011).

33. X. Jiao, J. Yin, X. Xu, J. Wang, Y. Liu, S. Xiong, Q. Zhang, and J. Song, *Advanced Functional Materials*, **31**, 2005699 (2021).

34. A. Magasinski, B. Zdyrko, I. Kovalenko, B. Hertzberg, R. Burtovyy, C. F. Huebner, T. F. Fuller, I. Luzinov, and G. Yushin, *ACS Applied Materials & Interfaces*, **2**, 3004 (2010).

35. S. Jiang, B. Hu, Z. Shi, W. Chen, Z. Zhang, and L. Zhang, *Advanced Functional Materials*, **30**, 1908558 (2020).

36. Z.-J. Han, K. Yamagiwa, N. Yabuuchi, J.-Y. Son, Y.-T. Cui, H. Oji, A. Kogure, T. Harada, S. Ishikawa, Y. Aoki, and S. Komaba, *Physical Chemistry Chemical Physics*, **17**, 3783 (2015).

37. L. Han, T. Liu, O. Sheng, Y. Liu, Y. Wang, J. Nai, L. Zhang, and X. Tao, *ACS Applied Materials & Interfaces*, **13**, 45139 (2021).

38. S. E. Trask, B. J. Polzin, J. Kubal, W. Lu, and A. N. Jansen, *ECS Meeting Abstracts*, **MA2015-01**, 301 (2015).

39. M.-T. F. Rodrigues, S. Rajendran, S. E. Trask, A. R. Dunlop, A. Singh, J. M. Allen, P. J. Weddle, J. I. Preimesberger, J. Coyle, A. M. Colclasure, Z. Yang, B. J. Ingram, and A. N. Jansen, *ACS Applied Energy Materials*, **6**, 9243 (2023).

40. C. Ye, M. Liu, X. Zhang, Q. Tong, M. Zhu, and J. Weng, *Journal of The Electrochemical Society*, **168**, 100519 (2021).

41. J. Choi, K. Kim, J. Jeong, K. Y. Cho, M.-H. Ryou, and Y. M. Lee, *ACS Applied Materials & Interfaces*, **7**, 14851 (2015).

42. B. N. Wilkes, Z. L. Brown, L. J. Krause, M. Triemert, and M. N. Obrovac, *Journal of The Electrochemical Society*, **163**, A364 (2016).

43. K. P. C. Yao, J. S. Okasinski, K. Kalaga, I. A. Shkrob, and D. P. Abraham, *Energy & Environmental Science*, **12**, 656 (2019).

44. D. P. Finegan, A. Quinn, D. S. Wragg, A. M. Colclasure, X. Lu, C. Tan, T. M. M. Heenan, R. Jervis, D. J. L. Brett, S. Das, T. Gao, D. A. Cogswell, M. Z. Bazant, M. Di Michiel, S. Checchia, P. R. Shearing, and K. Smith, *Energy & Environmental Science*, **13**, 2570 (2020).

45. J. Li, J. Fleetwood, W. B. Hawley, and W. Kays, *Chemical Reviews*, **122**, 903 (2022).

46. J. Kumberg, M. Müller, R. Diehm, S. Spiegel, C. Wachsmann, W. Bauer, P. Scharfer, and W. Schabel, *Energy Technology*, **7**, 1900722 (2019).

47. M.-T. F. Rodrigues, *Journal of The Electrochemical Society*, **169**, 080524 (2022).

48. B. R. Long, S. G. Rinaldo, K. G. Gallagher, D. W. Dees, S. E. Trask, B. J. Polzin, A. N. Jansen, D. P. Abraham, I. Bloom, J. Bareño, and J. R. Croy, *Journal of The Electrochemical Society*, **163**, A2999 (2016).

49. M.-T. F. Rodrigues, *Journal of The Electrochemical Society*, **169**, 110514 (2022).

50. S.-H. Kang, W.-S. Yoon, K.-W. Nam, X.-Q. Yang, and D. P. Abraham, *Journal of Materials Science*, **43**, 4701 (2008).

51. K. Schönherr, M. Pöthe, B. Schumm, H. Althues, C. Leyens, and S. Kaskel, *Batteries*, **9**, 53 (2023).

52. C. Yang, H. Ma, R. Yuan, K. Wang, K. Liu, Y. Long, F. Xu, L. Li, H. Zhang, Y. Zhang, X. Li, and H. Wu, *Nature Energy*, **8**, 703 (2023).

53. A. L. Musgrove, K. L. Browning, R. L. Sacci, A. Ullman, H. M. Meyer, III, K. Musgrove, J. Quinn, S. Möller, M. Finsterbusch, and G. M. Veith, *Energy & Fuels*, **39**, 4968 (2025).

54. K. Kalaga, M.-T. F. Rodrigues, J. Bareño, I. A. Shkrob, and D. P. Abraham, *Journal of Power Sources*, **438**, 227033 (2019).

55. J. D. McBrayer, N. B. Schorr, M. N. Lam, M. L. Meyerson, K. L. Harrison, and S. D. Minteer, *ACS Applied Materials & Interfaces*, **16**, 19663 (2024).

56. K. Zhang, Y. Zhang, J. Zhou, Y. Li, B. Zheng, F. Yang, and Y. Kai, *Journal of Energy Storage*, **32**, 101765 (2020).

57. M. Wetjen, S. Solchenbach, D. Pritzl, J. Hou, V. Tileli, and H. A. Gasteiger, *Journal of The Electrochemical Society*, **165**, A1503 (2018).

58. T. Waldmann, B.-I. Hogg, and M. Wohlfahrt-Mehrens, *Journal of Power Sources*, **384**, 107 (2018).

59. E. F. Woods, D. Wu, L. A. Robertson, H. Liu, B. Key, J. T. Vaughey, and Z. Zhang, *ACS Applied Energy Materials*, **7**, 8294 (2024).

60. J. C. Burns, R. Petibon, K. J. Nelson, N. N. Sinha, A. Kassam, B. M. Way, and J. R. Dahn, *Journal of The Electrochemical Society*, **160**, A1668 (2013).

61. J. Cannarella and C. B. Arnold, *Journal of Power Sources*, **245**, 745 (2014).

62. J. Brehm, M. Altmann, L. Mehlsam, P. Kotter, and A. Jossen, *Journal of The Electrochemical Society*, **172**, 070506 (2025).

63. S. Friedrich, S. Helmer, L. Reuter, J. L. S. Dickmanns, A. Durdel, and A. Jossen, *Journal of The Electrochemical Society*, **171**, 090503 (2024).

64. V. Müller, R.-G. Scurtu, K. Richter, T. Waldmann, M. Memm, M. A. Danzer, and M. Wohlfahrt-Mehrens, *Journal of The Electrochemical Society*, **166**, A3796 (2019).

65. S. Friedrich, S. Stojecevic, P. Rapp, S. Helmer, M. Bock, A. Durdel, H. A. Gasteiger, and A. Jossen, *Journal of The Electrochemical Society*, **171**, 050540 (2024).

66. U.S. Department of Energy Vehicle Technologies Program 2020 United States Advanced Battery Consortium Battery Test Manual For Electric Vehicles, Revision 3.1, in.

67. M. Naumann, M. Schimpe, P. Keil, H. C. Hesse, and A. Jossen, *Journal of Energy Storage*, **17**, 153 (2018).

68. K. Bischof, M. Flügel, M. Hölzle, M. Wohlfahrt-Mehrens, and T. Waldmann, *Journal of The Electrochemical Society*, **171**, 010510 (2024).

69. P. Gasper, N. Sunderlin, N. Dunlap, P. Walker, D. P. Finegan, K. Smith, and F. Thakkar, *Journal of Power Sources*, **604**, 234494 (2024).

70. A. Devie, M. Dubarry, and B. Y. Liaw, *Journal of The Electrochemical Society*, **162**, A1033 (2015).

71. D. Salpekar, J. I. Preimesberger, M.-T. F. Rodrigues, A. Singh, A. Verma, S. E. Trask, A. Colclasure, J. Coyle, A. N. Jansen, D. P. Abraham, and W. Lu, *Journal of Power Sources*, **655**, 237853 (2025).

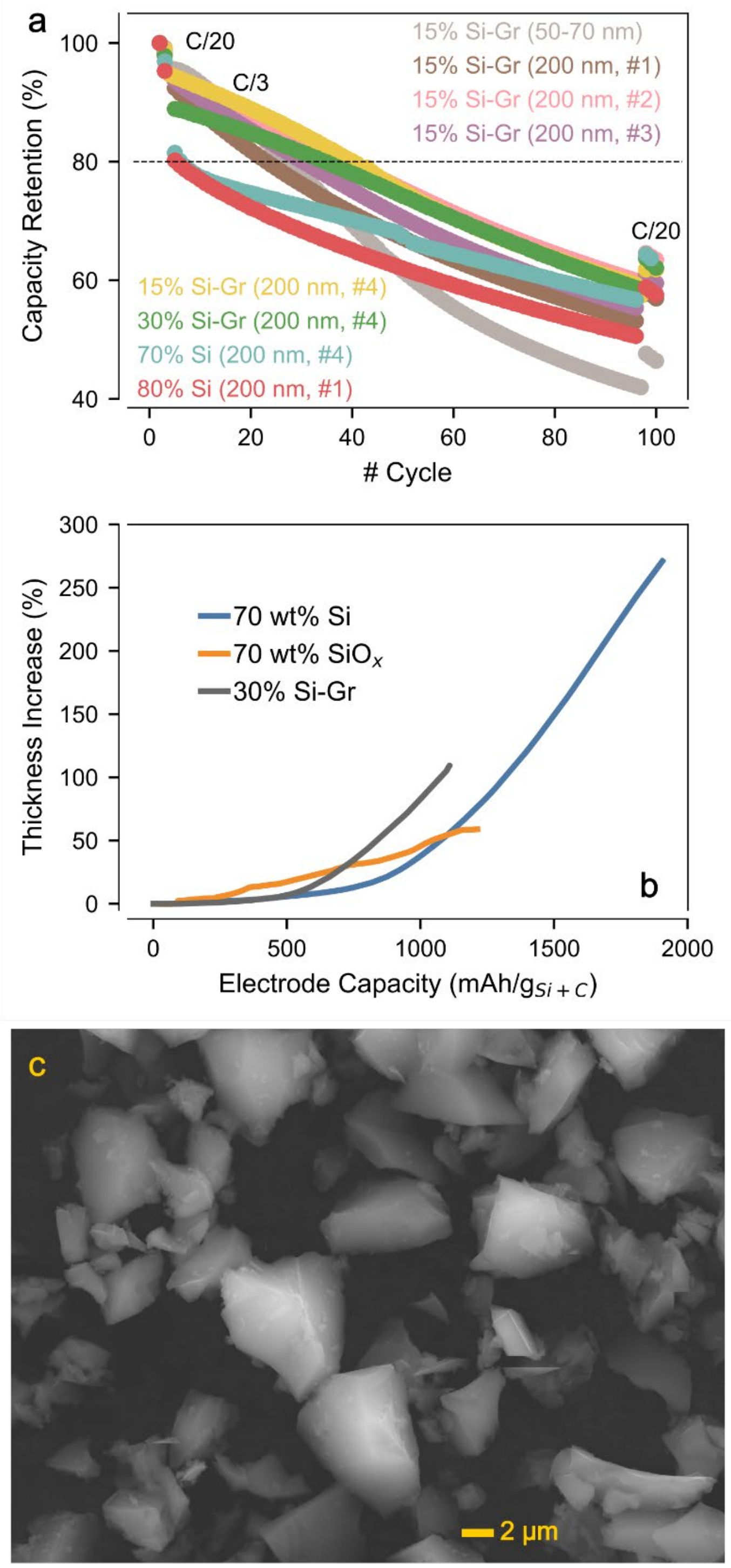

Figure 1. Active materials. a) Cycling data of full-cells containing NMC532 cathodes vs various Si-based anodes, using Gen2 + 10 wt% FEC as the electrolyte. Experiments with Si nanoparticles led to rapid capacity fade. Four different types of ~200 nm Si particles were tested (all from the same vendor but containing varying surface coatings); b) Thickness change measured through electrochemical dilatometry for the indicated electrodes. Capacity values were normalized to those measured in coin cells vs Li metal.

Undulations in the $SiO_x$ were caused by temperature variations during the test. All electrodes used LiPAA as binder; c) Scanning electron micrograph of $SiO_x$, showing micron-sized particles.

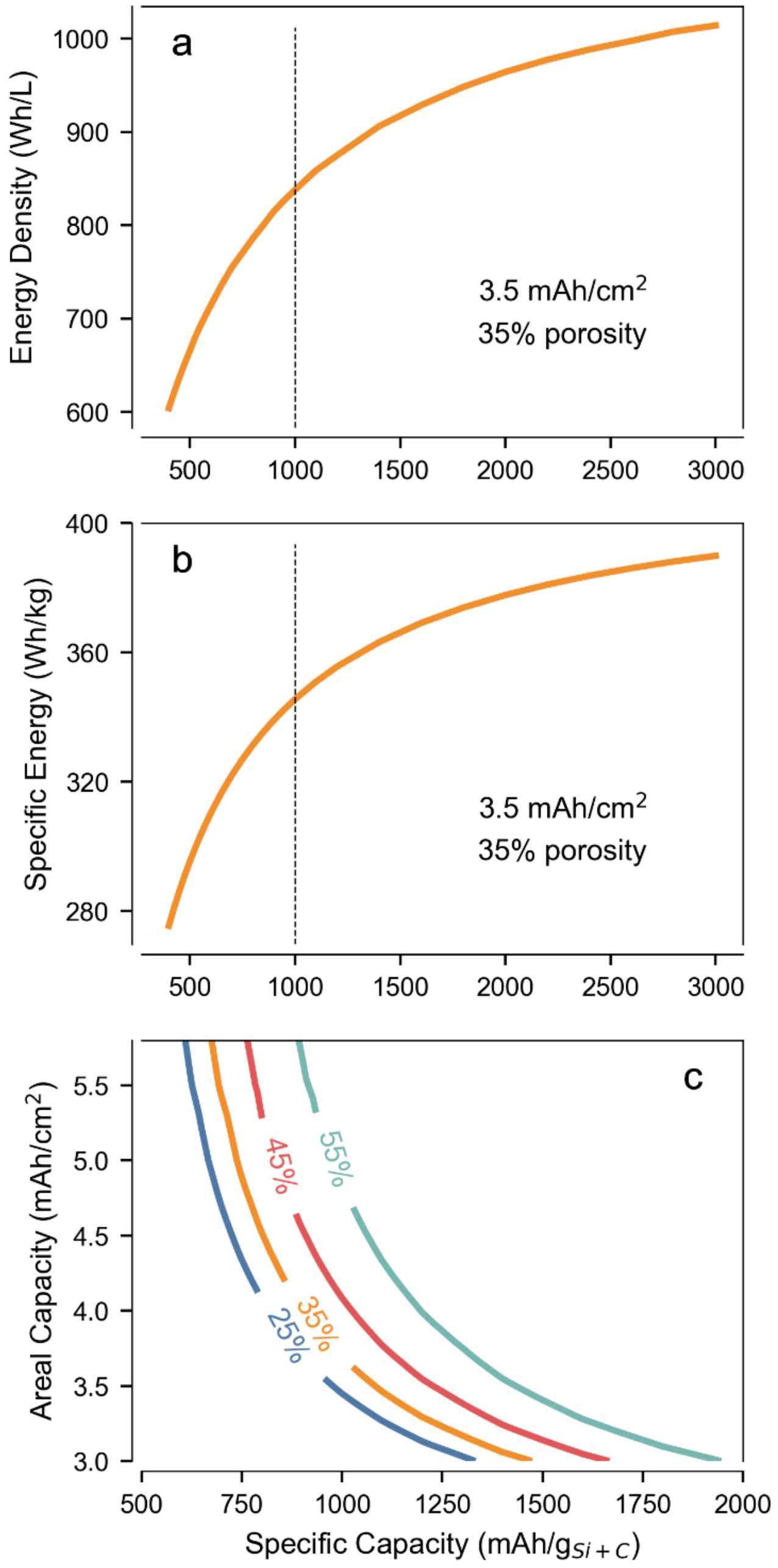

Figure 2. Performance projection of Si-containing electrodes vs NMC811 using BatPaC. a) Energy density and b) specific energy as a function of the specific capacity assumed for the active material at the indicated anode porosity and areal loading; c) Contour plots showing the combinations of anode areal and specific capacities that result in 350 Wh/kg on a cell level. Different curves assume different anode porosities.

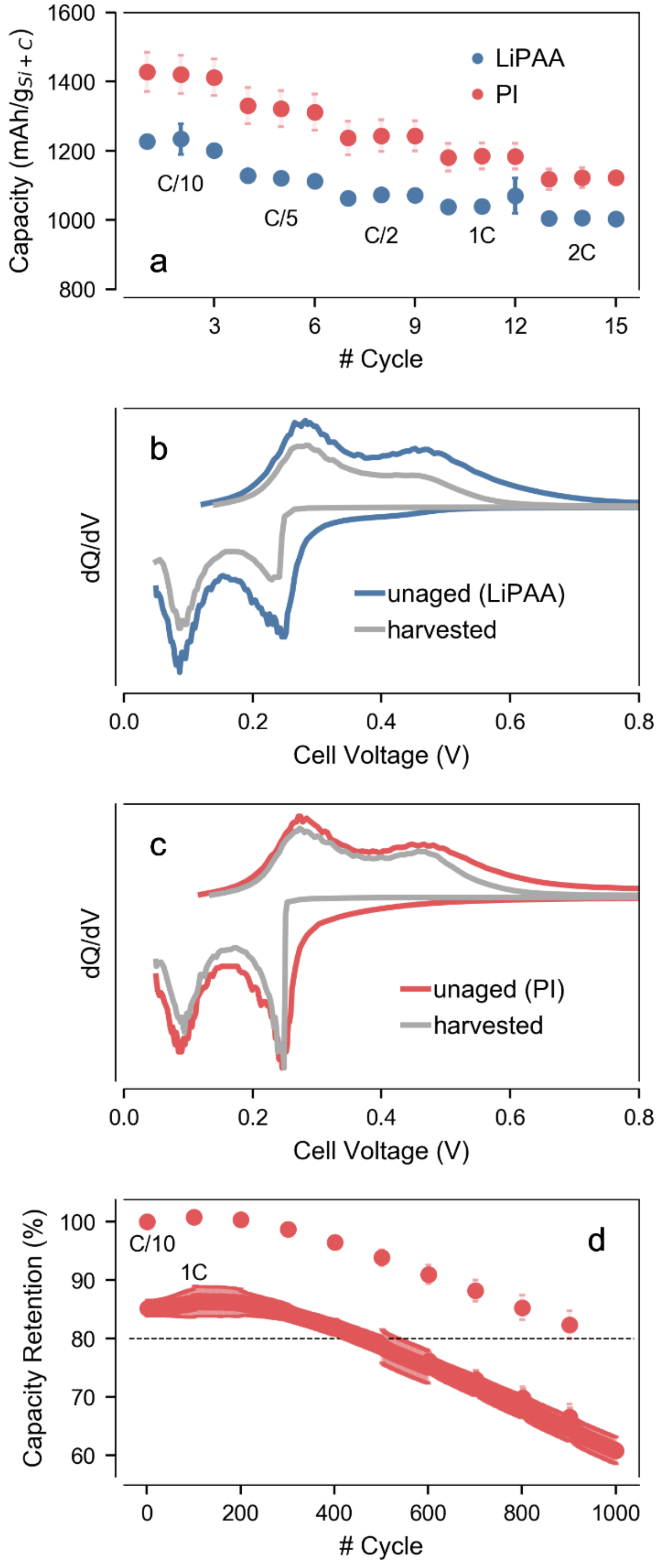

Figure 3. Testing of $SiO_x$ with two binders. a) Delithiation capacities from rate test of $SiO_x$ half-cells; dQ/dV profiles vs Li for cells after formation (unaged) and after aging (harvested) for $SiO_x$ electrodes with LiPAA (b) or polyimide (c) as binder; d) Cycling in coin cells of prelithiated $SiO_x$ vs NMC811 using Gen2 + 3wt% FEC as the electrolyte (pore volume factor = 4.5), showing that the electrode pair can achieve long life. Error bars indicate one standard deviation.

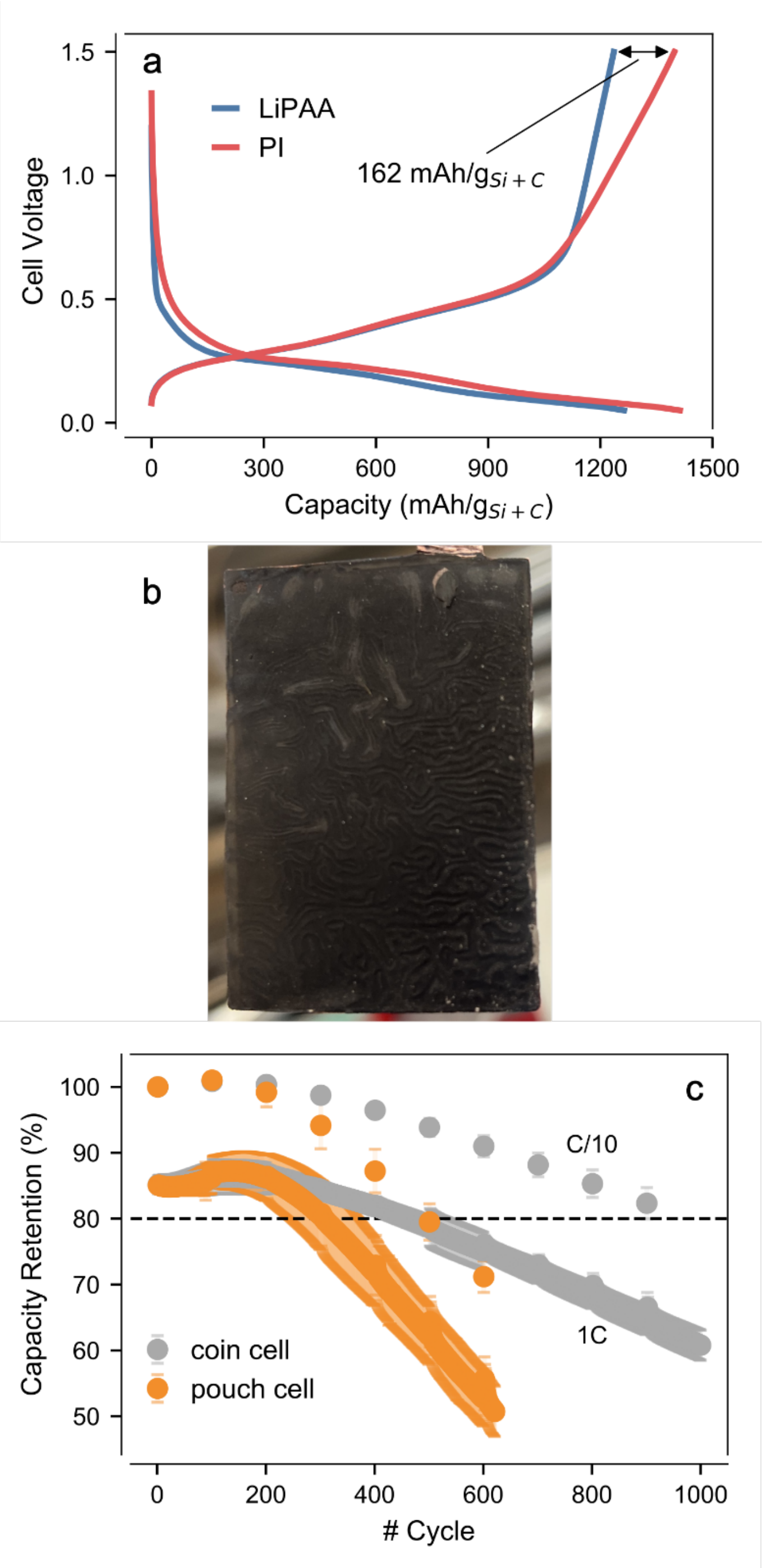

Figure 4. Challenges with the polyimide binder. a) Although electrodes with the PI exhibit higher capacity, these gains are due to a partially reversible redox reaction of the polymer itself rather than to improved utilization of the $SiO_x$; b) SiOx anode (PI binder) harvested from a pouch cell (14.1 $cm^2$; data in panel c). The superb adhesion strength of the PI leads to rippling of both electrode coating and Cu foil, and the resulting defects facilitate Li plating along the ridges; c) SiOx/PI electrodes tested in full-cells vs NMC811 in two different cell formats (both with pore volume factor = 4.5). Rippling does not happen in small coin cells but becomes possible in large pouch cells, where cell life becomes limited by the occurrence of Li plating. Error bars indicate one standard deviation.

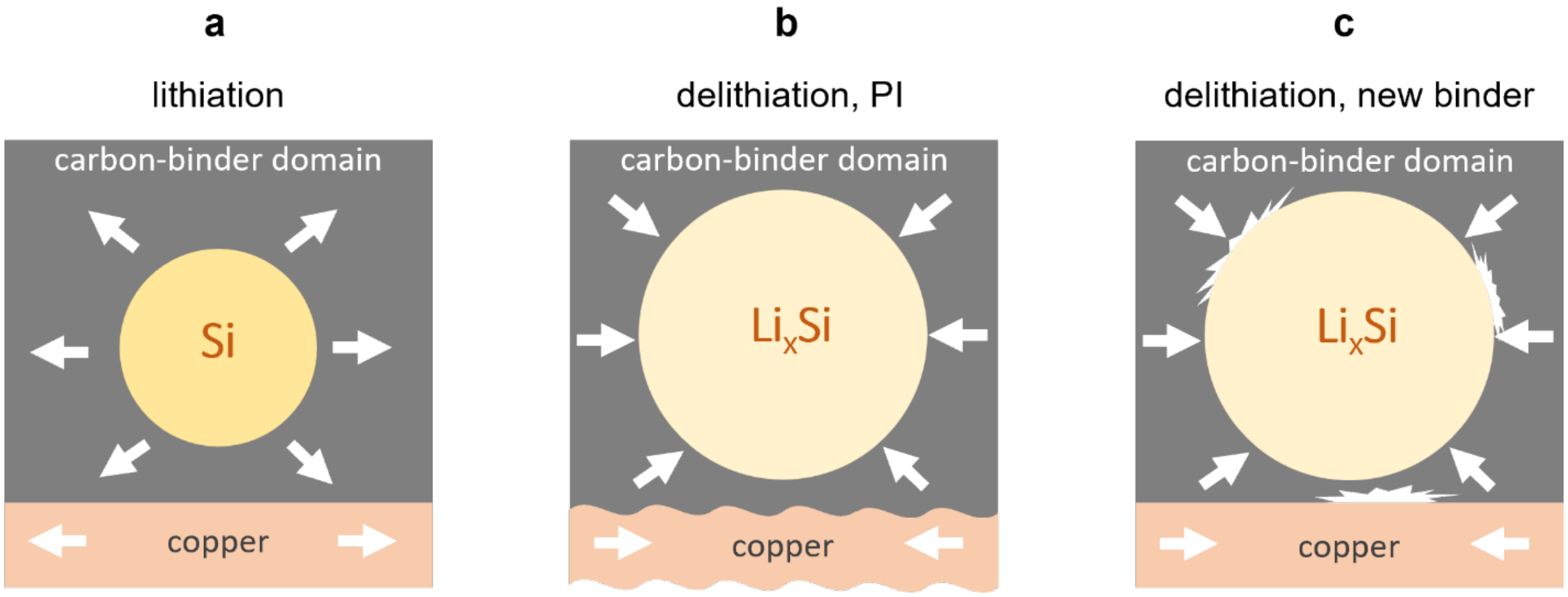

Figure 5. Schematic representation of the formation of ripples during cycling. a) During anode lithiation, a sufficiently strong binder will result in stresses from Si dilation being transferred to the Cu foil. If the stress is too high, plastic deformation of the foil may ensue; b) Silicon contraction during delithiation creates compressive stress in the Cu. For electrode dimensions larger than a critical size, this inward pull will result in wrinkles; c) A binder with suitable adhesion strength will lead to small, localized interfacial debonding, partially releasing cycling-related stresses and thus averting plastic deformation.

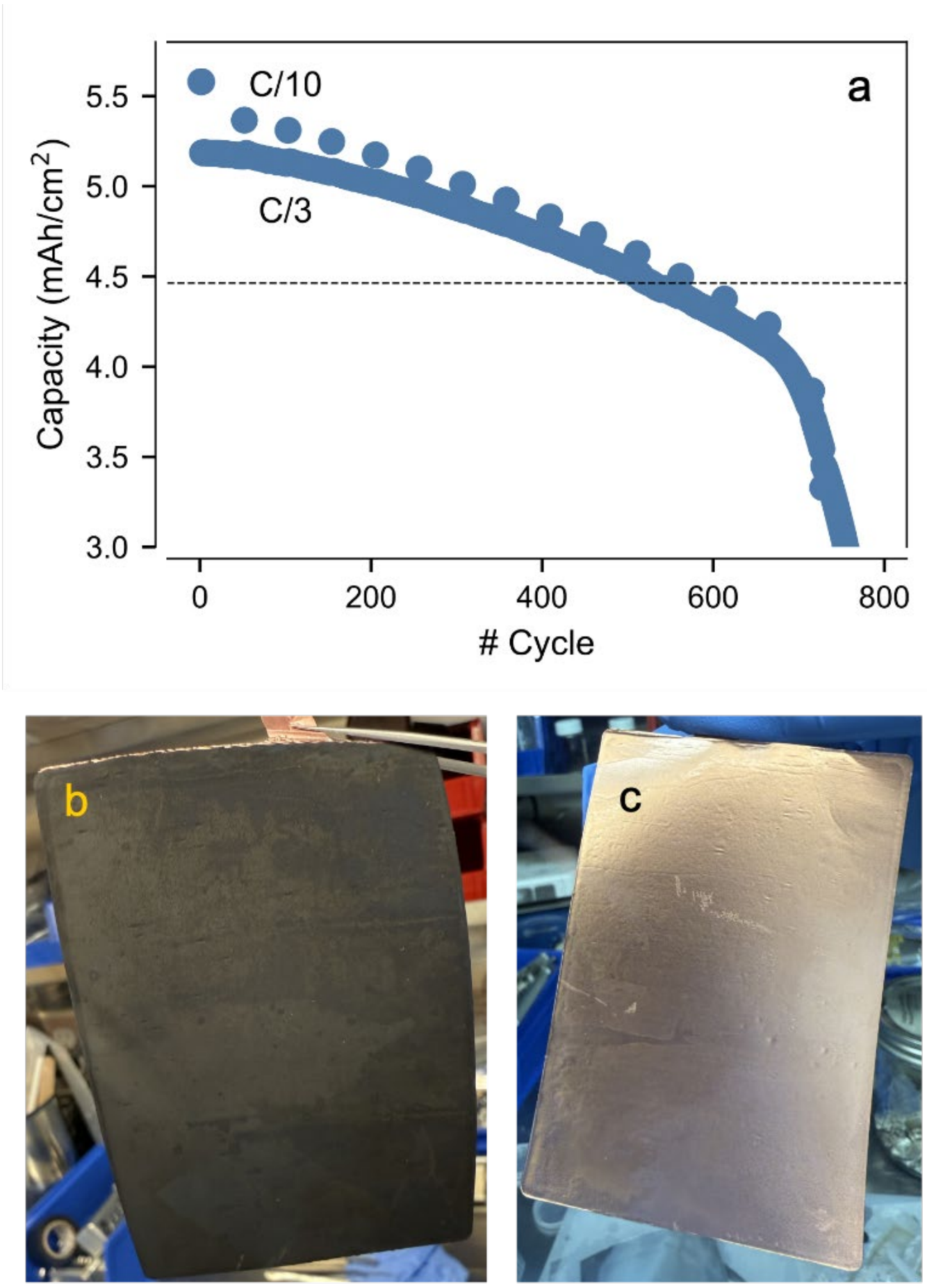

Figure 6. High-loading $SiO_x$ electrodes (> 5.5 mAh/cm$^2$) without wrinkling. a) Cycling data of 46.3 cm$^2$ pouch cells containing NMC811 cathode vs an anode using a polyacrylate binder and using Gen2 + 3 wt% FEC as the electrolyte (pore volume factor = 4). The horizontal dashed line indicates 80% retention of the initial C/10 capacity; Cell teardown at the end of testing showed no signs of Li plating (b) or rippling (c), suggesting that rollover failure was caused by electrolyte depletion.

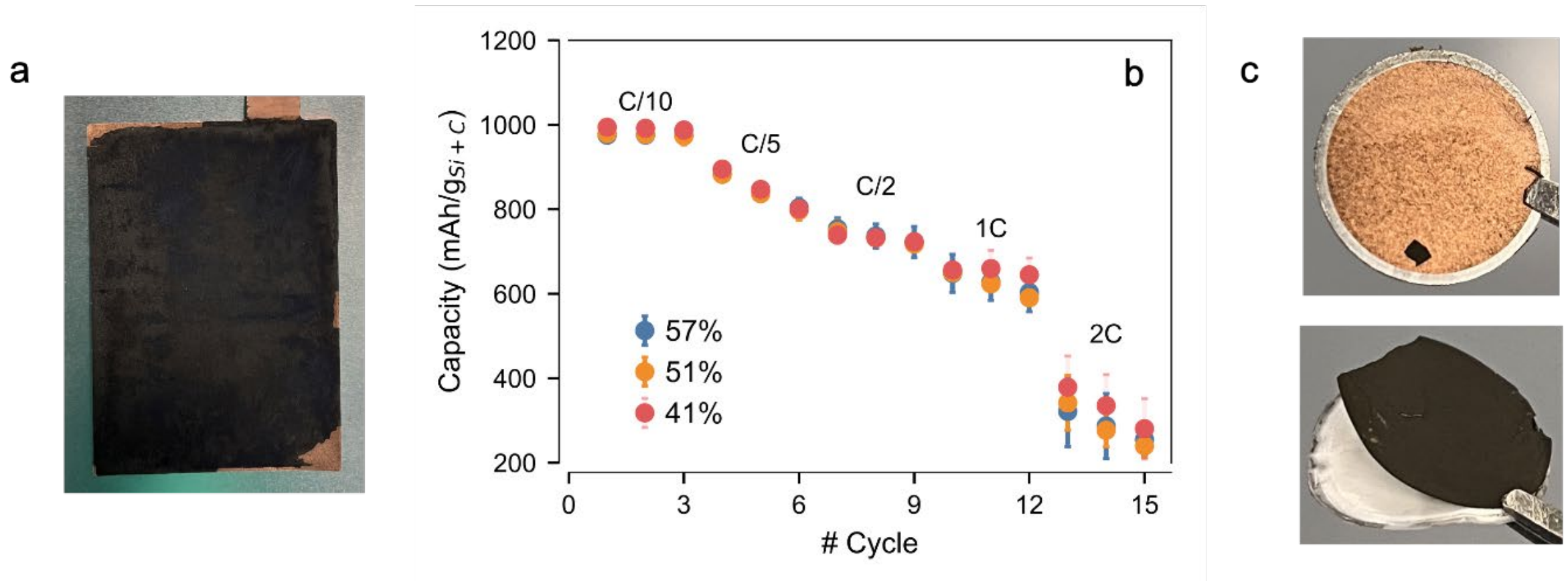

Figure 7. Initial challenges with the polyacrylate binder. a) Anodes with > 5.5 mAh/cm$^2$ and 57% porosity showed occasional delamination along electrode edges when dismantling pouch cells after prelithiation; b) Rate test of $SiO_x$ half-cells with the anode calendered to different porosities; c) Observation from dismantling half-cells containing anodes with 41% porosity. At lower porosities, stresses from $SiO_x$ volume change led to complete delamination of the coating. Error bars indicate one standard deviation.

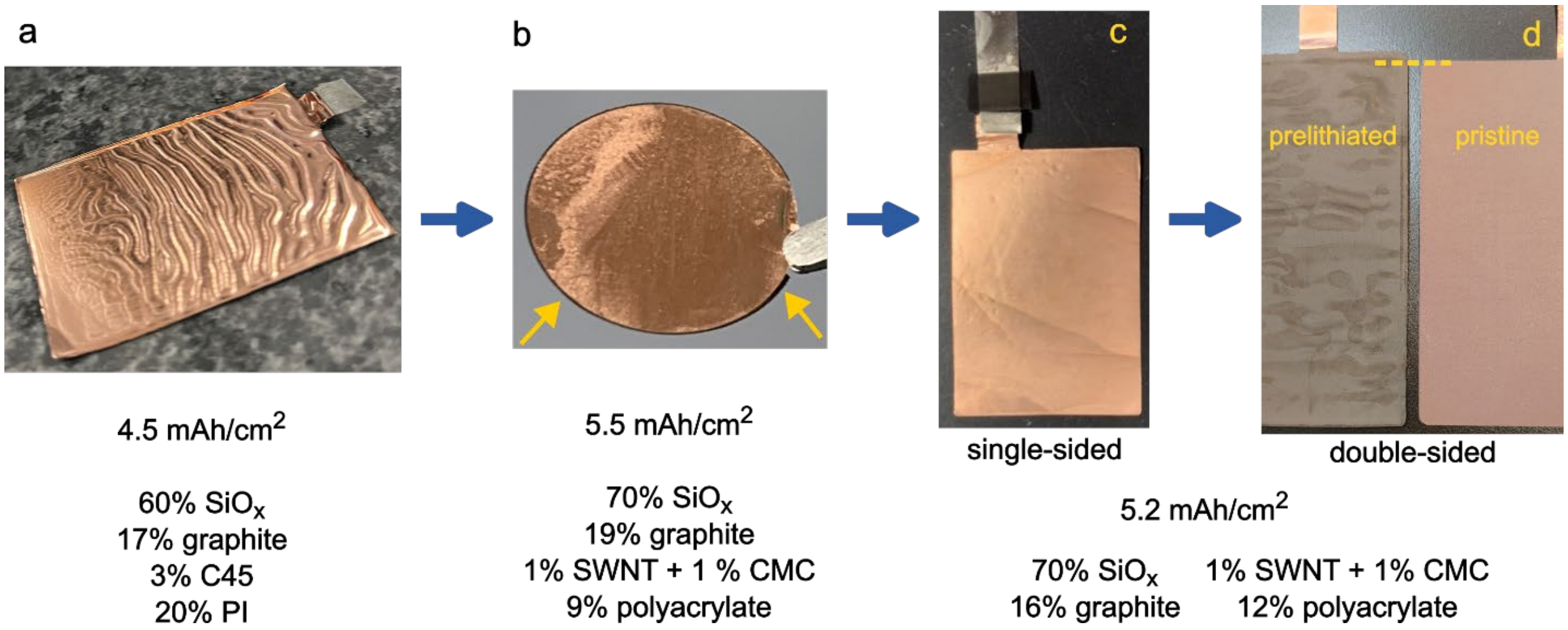

Figure 8. Evolution of electrode formulation in response to mechanical issues. a) The initial PI-based composition led to severe wrinkling of coating and foil; b) Adopting a polyacrylate binder mitigated wrinkling but led to the coating often expanding beyond the bounds of the foil (see yellow arrows), facilitating edge delamination when handling after prelithiation; Slightly increasing the polyacrylate content led to improved coating adhesion while still preventing wrinkling (c), but the addition of a second layer of coating at the electrode back-side resulted in plastic deformation of copper (d).

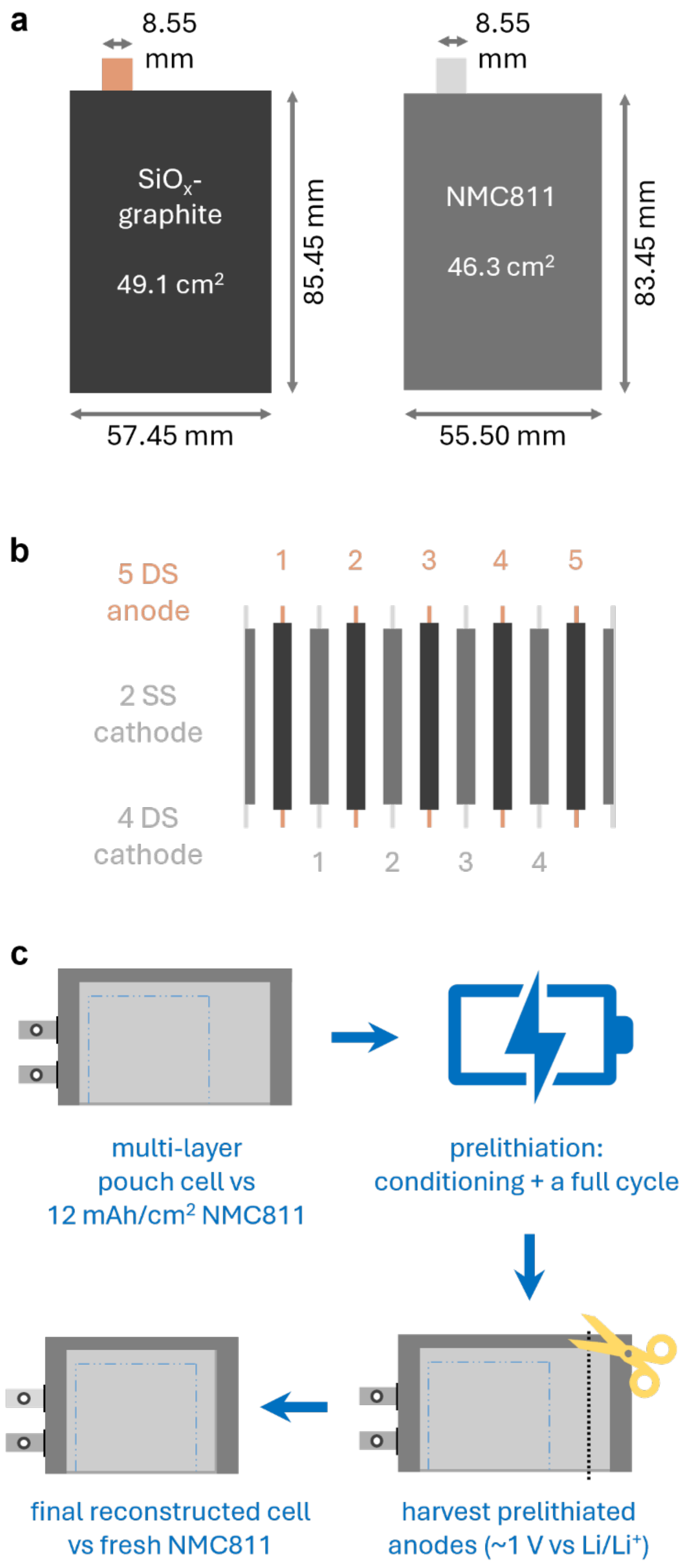

Figure 9. Building ~2 Ah multi-layer pouch cells. a) Electrode dimensions for a 46.3 $cm^2$ cell; b) Schematic of the stacking strategy, with all anode layers being double-sided (DS) and the extremes presenting single-sided (SS) cathodes; c) Prelithiation workflow, involving the assembly of a cell with thick "sacrificial" NMC811, anode harvesting and reutilization in the final reconstructed pouch cells.

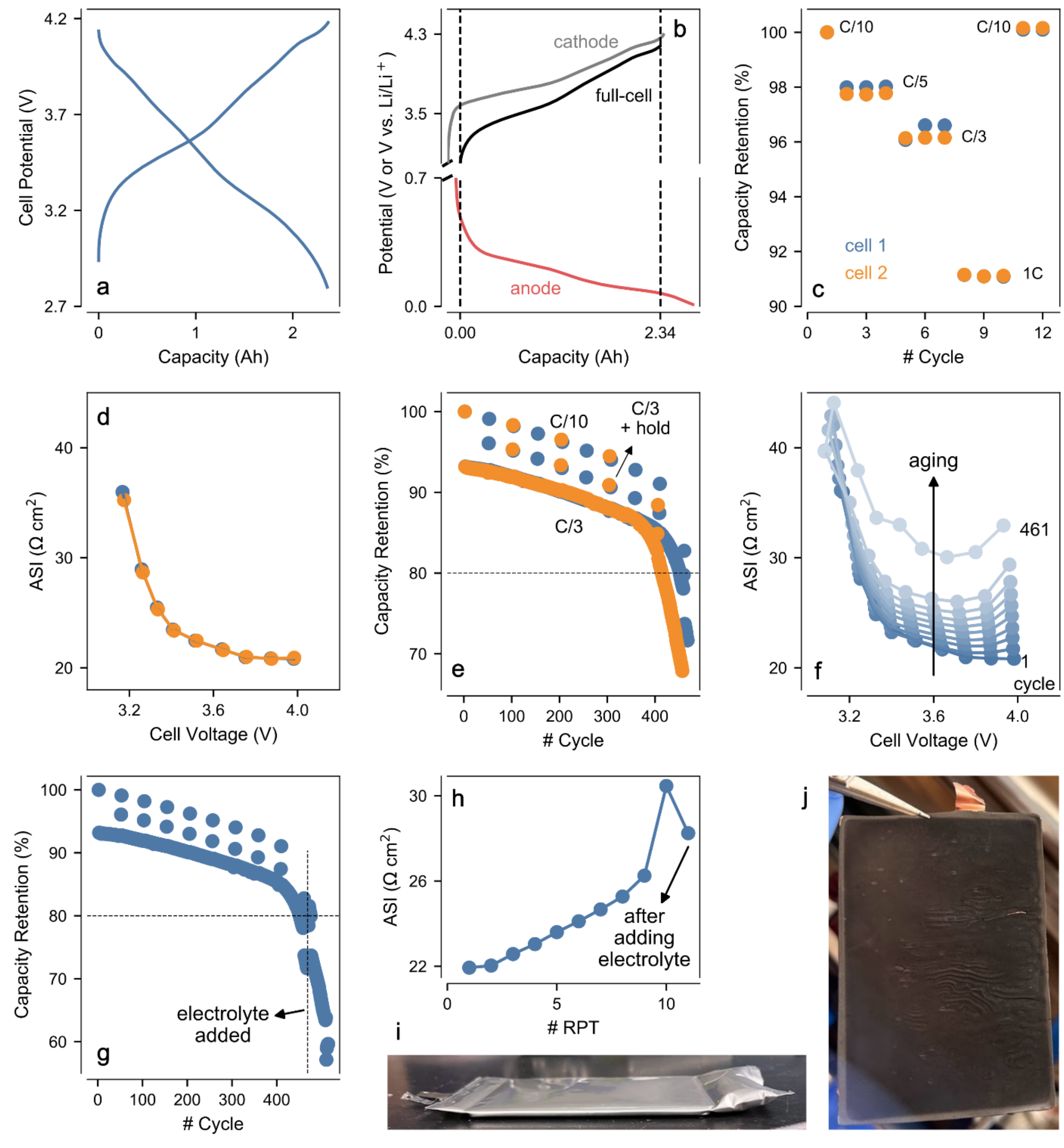

Figure 10. Multi-layer pouch cell testing with Gen2 + 3 wt% FEC as the electrolyte. a) C/10 charge and discharge profiles at the beginning of life, showing > 2.3 Ah; b) Approximate alignment of the voltage profiles of cathode and anode during cell charge obtained through differential voltage analysis. Portions of the profile to the left of zero and to the right of 2.34 Ah are not actively utilized in that cycle; c) Rate test, showing capacities at the indicated rates with respect to the initial C/10 capacity of the cells. d) Initial area-specific impedance (ASI) measured through discharge pulses; e) Cycling data, exhibiting consistent rollover after ~400 cycles; f) ASI measured during testing of cell 1, exhibiting a sudden increase that coincided with the rollover in panel e; Adding extra electrolyte to cell 1 after the rollover did not change the rate of fade (g), but led to a drastic decrease in impedance (h); i) cell 1 at the end of testing, with visible gassing; j) Example of anode from cell 1 after teardown, exhibiting Li plating at the ridges of the deformed electrode (the grayish band at the ends is the overhang). The color code of panel e applies to all other panels. These cells had a pore volume factor = 2.15.

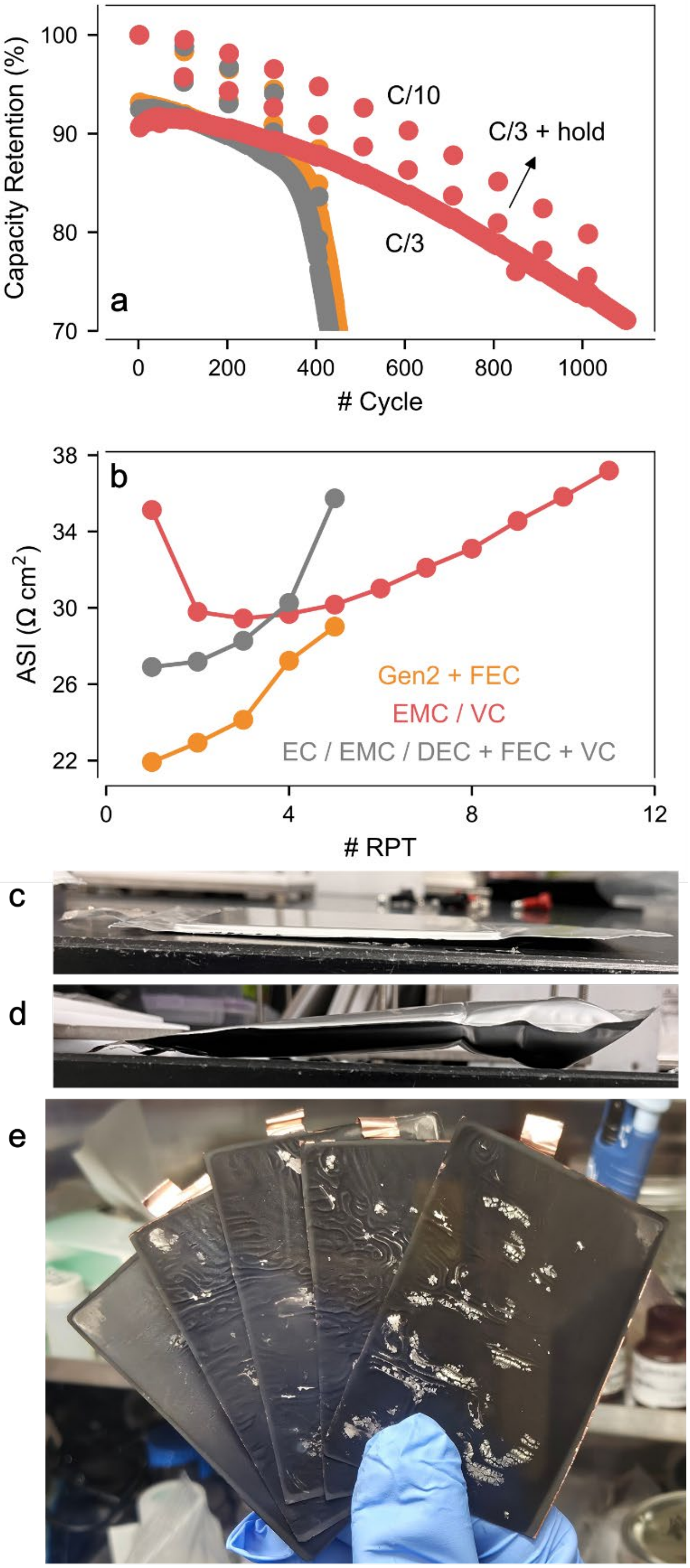

Figure 11. Multi-layer pouch cells containing other electrolyte formulations. a) Cycling data, showing that only the EC-free, EMC/VC composition could avoid rollover; b) Cell impedances interpolated at 3.6 V at various reference performance tests (RPTs). The initial ASI drop of the EMC/VC was consistently observed in all cell formats we tested; Photographs of cells at the end of testing with the EMC/VC

electrolyte (c) and with the EC/EMC/DEC composition (d). Remarkably, the former showed negligible gassing after 1,000 cycles; e) Anodes from the EMC/VC cell, showing that long life and no gassing were achieved despite significant Li plating. The color code in panel b also applies to panel b. See text for full composition of electrolytes. These cells had a pore volume factor = 2.15.

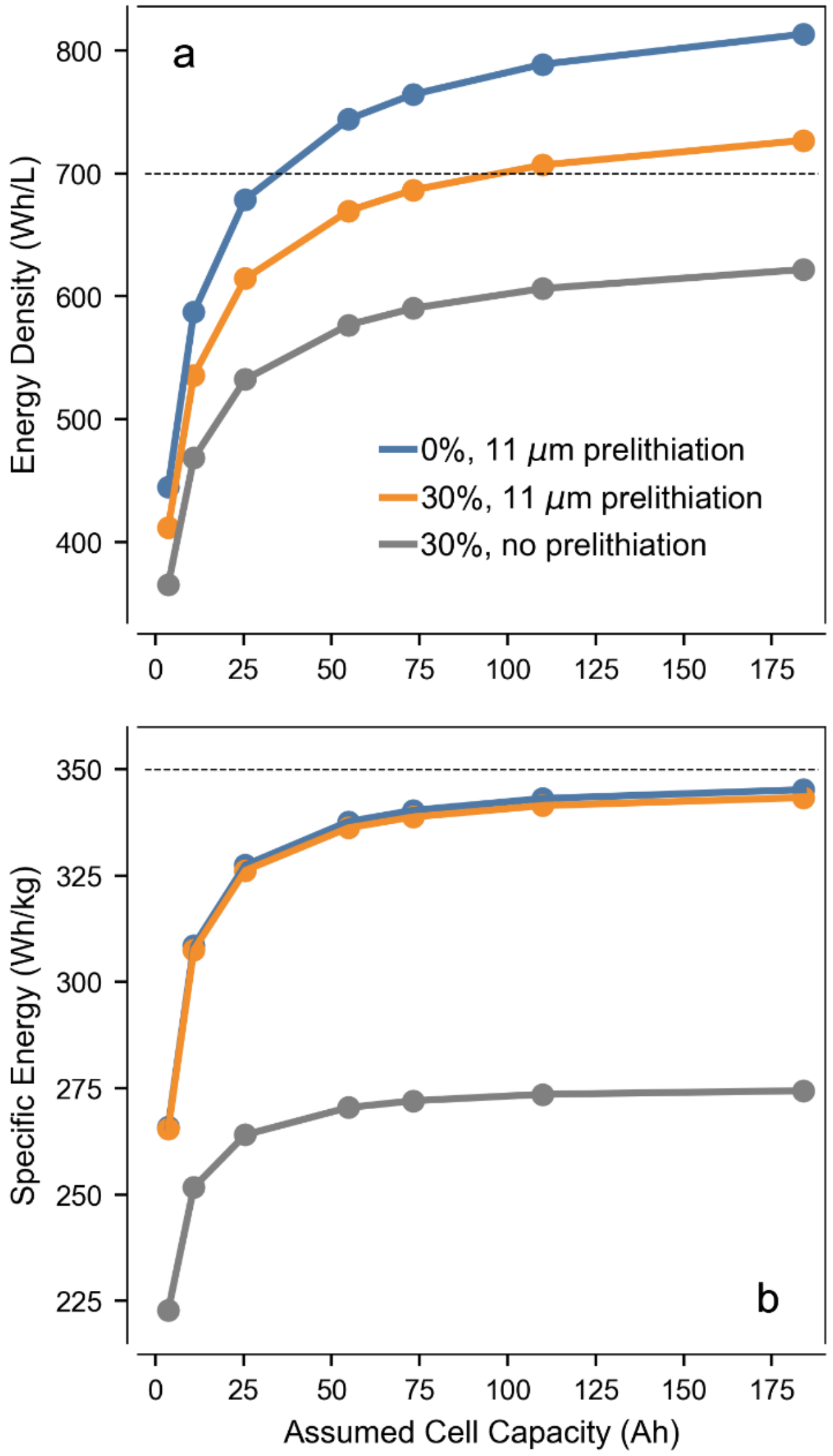

Figure 12. Projected energy metrics for the final anode formulation using BatPaC. a) Energy density; b) Specific energy. Anode properties were the same as exhibited by the tested $SiO_x$ electrodes. Not prelithiating the anode would require much thicker cathodes, decreasing energy metrics (gray). Accommodating the thickness dilation of the anode in the cell design leads to a decrease in Wh/L but negligible changes in Wh/kg. The horizontal dashed lines mark our target metrics. The color code of panel a also applies to panel b.

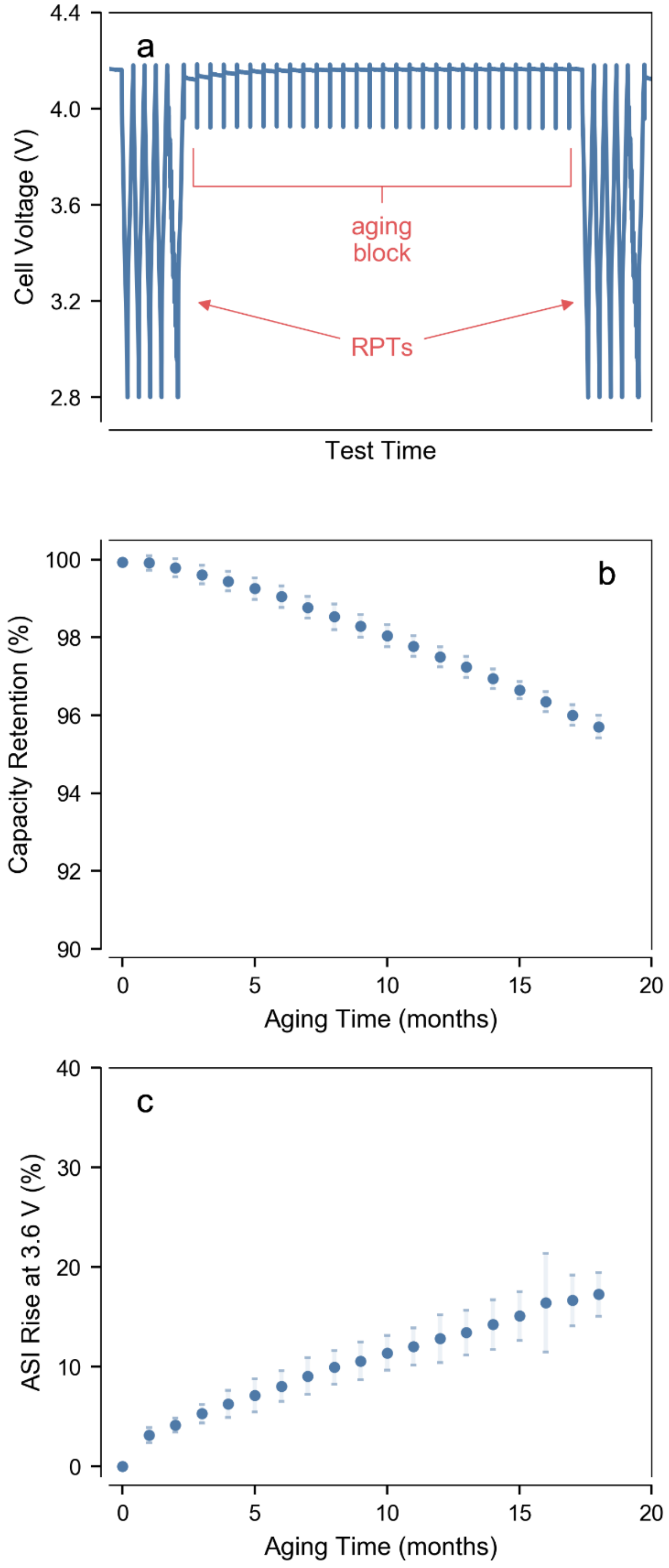

Figure 13. Calendar aging experiments in 14.1 $cm^2$ single-layer pouch cells using Gen2 + 3 wt% FEC as electrolyte. a) Schematic representation of test protocol, in which month-long aging periods are spaced by RPTs. Cells remain at open circuit during aging, with daily charges to 100% SOC to counter voltage drifts from self-discharge; b) Capacity retention as measured from the third C/10 cycle of each RPT,

showing remarkably slow fade; c) Impedance interpolated at 3.6 V as a function of total aging time. These cells had a pore volume factor = 4.4. Error bars indicate one standard deviation.

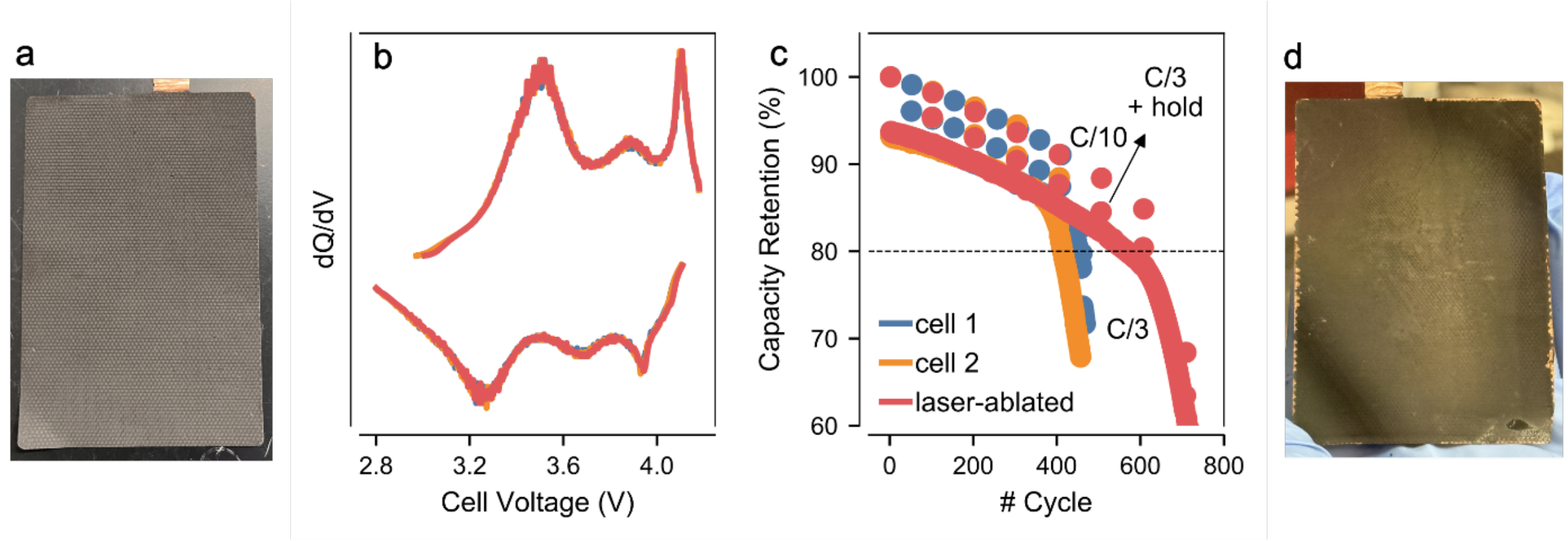

Figure 14. Mitigating rippling using laser ablation. a) Photograph of a double-sided laser ablated $SiO_x$ electrode after prelithiation, showing that trenches are preserved and no wrinkling is observed; b) dQ/dV profiles at the beginning of life demonstrate that bilayer cells containing the ablated anode are indistinguishable from the multi-layer pouch cells; c) Cycling data, showing that laser ablation can prolong cell life; d) Photograph of laser-ablated anode after testing, showing no signs of rippling or Li plating. The color code of panel c also applies to panel b. These cells had a pore volume factor = 2.15.

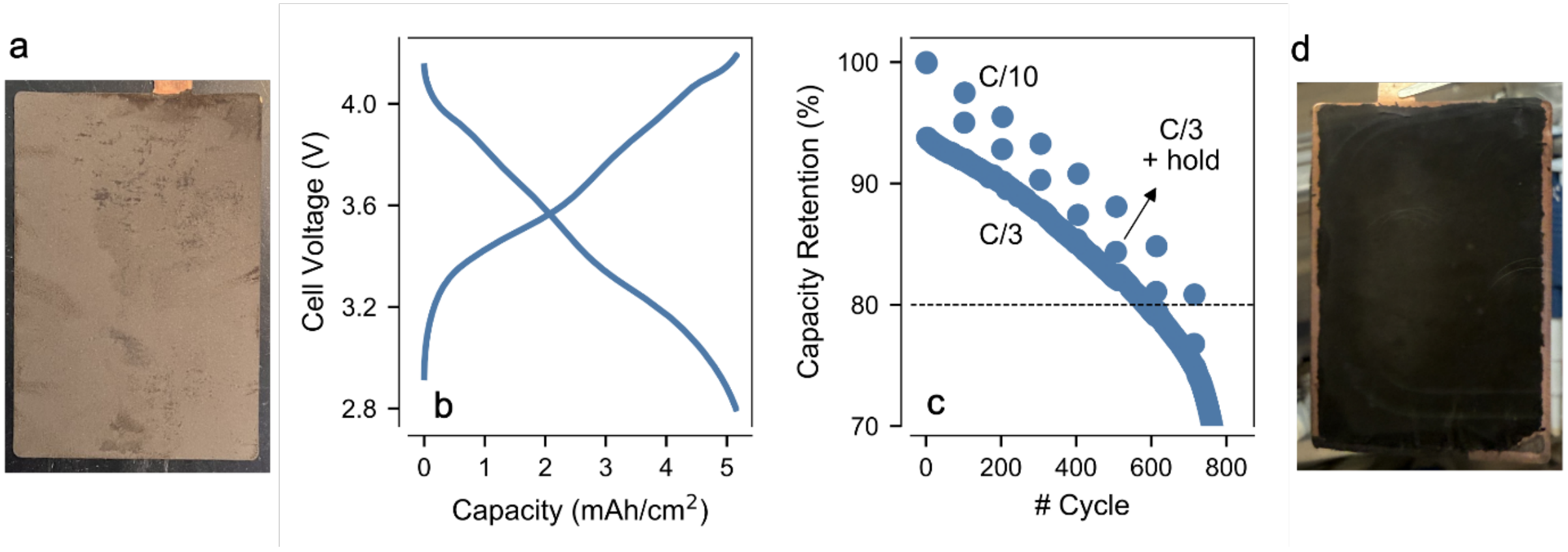

Figure 15. Exploring improved electrode formulations. a) Photograph of double-sided electrode after prelithiation, showing no deformation of coating or current collector; b) Voltage profiles at C/10 at the beginning of life, indicating that > 5 mAh/cm$^2$ per side of coating is achieved; c) Cycling data of bilayer pouch cell with Gen2 + 3 wt% FEC (pore volume factor = 3), showing ~700 cycles of life; d) Photograph of anode after testing, showing no signs of rippling or Li plating.